\documentclass[aps,twocolumn,amsmath,amssymb,preprintnumbers]{revtex4}
\usepackage{amsmath} \usepackage{amsfonts} \usepackage{amssymb}
\usepackage{bbm}
\usepackage{epsfig}
\usepackage{graphics}
\usepackage{graphicx}
\newcommand{\be}{\begin{equation}}
\newcommand{\ee}{\end{equation}}
\newcommand{\ba}{\begin{eqnarray}}
\newcommand{\ea}{\end{eqnarray}}
\newcommand{\nn}{\nonumber}
\newcommand{\kr}{\rangle}
\newcommand{\kl}{\langle}

\newcommand{\cD}{{\cal D}}

\newcommand{\cN}{{\cal N}}
\newcommand{\cA}{{\cal A}}
\newcommand{\cK}{{\cal K}}
\newcommand{\cB}{{\cal B}}
\newcommand{\cP}{{\cal P}}
\newcommand{\cG}{{\cal G}}
\newcommand{\cR}{{\cal R}}
\newcommand{\mn}{\mu\nu}

\begin{document}

\title[ ]{Quantum field theory from classical statistics}

\author{C. Wetterich}
\affiliation{Institut  f\"ur Theoretische Physik\\
Universit\"at Heidelberg\\
Philosophenweg 16, D-69120 Heidelberg}

\begin{abstract}
An Ising-type classical statistical model is shown to describe quantum fermions. For a suitable time-evolution law for the probability distribution of the Ising-spins our model describes a quantum field theory for Dirac spinors in external electromagnetic fields, corresponding to a mean field approximation to quantum electrodynamics. All quantum features for the motion of an arbitrary number of electrons and positrons, including the characteristic interference effects for two-fermion states, are described by the classical statistical model. For one-particle states in the non-relativistic approximation we derive the Schr\"odinger equation for a particle in a potential from the time evolution law for the probability distribution of the Ising-spins. Thus all characteristic quantum features, as interference in a double slit experiment, tunneling or discrete energy levels for stationary states, are derived from a classical statistical ensemble. Concerning the particle-wave-duality of quantum mechanics, the discreteness of particles is traced back to the discreteness of occupation numbers or Ising-spins, while the continuity of the wave function reflects the continuity of the probability distribution for the Ising-spins.
\end{abstract}

\maketitle

\section{Introduction}
\label{Introduction}
Can the quantum-mechanical interference pattern in a double-slit experiment be described by a suitable time evolution of a classical statistical ensemble? Can this be extended to the quantum-interference of multi-fermion systems? Can the evolution law for the classical probabilities be causal and local in the sense that the probability distribution at time $t+\epsilon$ can be computed from the one at time $t$, and that the evolution of local properties of the distribution is only influenced by the distribution in a local neighborhood? We answer these questions with a clear ``yes''. While a general discussion how ``no-go-theorems'' as Bell's inequalities \cite{Bell,BS} or the Kochen-Specker theorem \cite{KS}  can be circumvented can be found in ref. \cite{GR}, we concentrate in this paper on the construction of a classical statistical model that can describe real physical situations. 

We will investigate an Ising-type model for discrete occupation numbers or Ising-spins on the sites of a lattice. Intuitively, if a particle is present on a lattice site $x$ the Ising-spin is up, and if no particle is present it is down. Since the associated occupation numbers only take the values one or zero, one sees already the close analogy to occupation numbers of fermionic multi-particle systems. The notion of presence or absence of particles can be extended to more general ground states, as a half-filled state. Now a flip of the Ising-spin at $x$ from down to up corresponds to a change from a situation with no particle at $x$ to one with a particle at $x$. More precisely, we will consider four or eight ``species'' of Ising-spins. These species correspond to the real degrees of freedom for Majorana or Dirac spinors. 

The state of the system or the classical statistical ensemble is characterized by the probability distribution for the possible configurations of Ising-spins. For the formulation of dynamics one has to postulate an evolution law which describes how this probability distribution changes in time. For the Ising-spins there is no notion of underlying continuous and deterministic classical dynamics for ``trajectories'' -each Ising-spin can only take two values. A differential evolution equation has therefore to be formulated on the level of probability distributions. This constitutes the fundamental law defining the dynamics. (It is the analogue of the Liouville equation for a statistical ensemble of Newtonian classical particles.) The only basic restrictions on the form of this law are that all probabilities have to remain positive and the sum of all probabilities must equal one for all times $t$. We propose a specific causal and local evolution law. With this evolution law our Ising-type model describes a quantum field theory for Dirac fermions in an arbitrary external electromagnetic field. The Schr\"odinger equation for a quantum particle in a potential follows for one-particle states in the non-relativistic approximation.

With our proposed evolution equation for a classical statistical ensemble of Ising-spins, the Schr\"odinger equation and all associated quantum phenomena as interference, tunneling, the uncertainty relation and discrete energy levels for stationary states, are implemented within classical statistics. We follow here the concept of ``probabilistic realism'' that there is one reality, but the fundamental laws are formulated as probabilistic laws. It seems not entirely excluded that our evolution law for the probability distribution could be produced by cellular automata \cite{TH}. This seems, however, rather difficult, and it is not needed for our purpose. 

Time evolution laws for classical statistical ensembles that account for the full dynamics of quantum particles have been found previously for more restricted settings. For a single particle in a potential a suitable evolution equation for the probability density in phase space has been proposed in ref. \cite{CWP}. It leads to the Schr\"odinger equation after ``coarse graining''. This ``quantum evolution equation'' modifies Liouville's equation. In ref. \cite{CWP} the different behavior of classical and quantum particles is traced back to the specific form of the evolution equation. A continuous interpolation between the ``classical'' and ``quantum'' evolution law can describe ``zwitters'' - particles whose properties interpolate continuously between classical and quantum particles. This earlier approach constitutes already a powerful demonstration that there is no difference in principle between classical statistics and quantum statistics since a continuous interpolation between both is possible. It also allows the formulation of consistent theories that are arbitrarily close to quantum mechanics, with a small parameter accounting for deviations that can be restricted by experiment. However, the generalization of this setting to a multi-particle situation is cumbersome. It is much simpler to base a realistic model directly on the analogy between Ising-spins and occupation numbers for fermions, as done in the present paper. Furthermore, the setting of ref. \cite{CWP} does not account explicitly for the discrete particle properties, while in the present paper they can be associated directly to the discrete Ising-spins. 

In ref. \cite{CWF}, \cite{CWES},  \cite{CWMSa} the map between Ising-type models and quantum fermions has been developed. In particular, ref. \cite{CWMSa} has already constructed a classical statistical ensemble for massless Majorana spinors in four dimensions. The present paper builds on these findings. It extends this formalism to include a mass term, to formulate Dirac spinors and to add the coupling to external electromagnetic fields. These steps are needed for a description of realistic situations. The discussion of one-particle states can now describe electrons in a potential. For a non-relativistic approximation we can derive the Schr\"odinger equation from the proposed evolution law for the classical statistical ensemble of Ising-spins. 

Our Ising-type model is formulated for a lattice of discrete points in space and we use discrete time steps. (The lattice formulation of the present paper differs in some aspects from ref. \cite{CWMSa}.) As long as the number of lattice points (in space) remains finite all mathematical operations are defined unambiguously and the model is fully regularized. The map between classical statistics for the Ising-type model and quantum fermions can be done on this level where no ambiguities are present. At the very end one may perform the continuum limits of vanishing lattice distance and vanishing size for the time steps. We formulate the model in a way such that the unitary time evolution is already realized for discrete time steps and a finite number of lattice points in space. This implies certain restrictions on the precise lattice formulation for which more details will be discussed in a separate paper.

Our paper is organized as follows: In sect. \ref{Probability distribution and wave function} we formulate the classical statistical ensemble for Ising-spins and we introduce the important concept of the ``classical wave function''. Up to signs this is the square root of the probability distribution. Time evolution laws preserving the positivity and normalization of the probability distribution can be most easily formulated in terms of the classical wave function. Since the normalization condition states that the classical wave function is a vector with unit length, we can formulate the normalization preserving evolution equations as rotations of a vector. We also employ the simple map between the classical wave function and a Grassmann wave function. It has been shown in refs. \cite{CWF}, \cite{CWMSa} that a linear time evolution of a Grassmann wave function can be mapped to a Grassmann functional integral and vice versa. This permits us to formulate the evolution law for the classical wave function and probability distribution directly in terms of the action of a Grassmann functional integral. This possibility is of great help for the implementation of the symmetries preserved by the evolution as, for example, the Lorentz symmetry for relativistic particles. 

In sect. \ref{Massless Majorana fermions} we present the lattice action for massless Majorana fermions in four dimensions. Since the classical wave function is a real object, it should be an element of a real Grassmann algebra. The continuum limit of the lattice action realizes Lorentz-symmetry. In sect. \ref{Grassmann wave function} we extract the time evolution of the Grassmann wave function from the real Grassmann functional integral defined in sect. \ref{Massless Majorana fermions}. It obtains by ``integrating'' out the Grassmann variables at past (or future) times. Sect. \ref{Timeevolution} shows that the time evolution is unitary. This holds already for discrete time steps and can therefore be extended in a straightforward way to the continuum limit of infinitesimal time steps. Sect. \ref{Timeevolution}  establishes the concrete evolution law for the classical wave function and therefore the probability distribution of the classical ensemble of Ising-spins. In sect. \ref{Observables} we briefly recapitulate how expectation values of observables in the classical statistical ensemble can be computed from the Grassmann wave function.

In sect. \ref{Particlestates} we turn to the notion of statistical states with a fixed number of fermions. We concentrate on the one particle state and show that it describes the propagation of a massless Majorana or Weyl fermion. This can be generalized to multi-fermion states. The equivalence between Majorana and Weyl spinors in four dimensions gives a first glance on the emergence of a complex structure which is discussed in sect. \ref{Complex structure}. The presence of a complex structure is characteristic for quantum mechanics where the ``physics of phases'' plays a crucial role. While single Majorana spinors are described by a real four-component quantum wave function, Weyl spinors are formulated in terms of an equivalent complex two-component quantum wave function. For Dirac spinors a different complex structure will be used and we therefore resume in sect. \ref{Complex structure} briefly the general features of complex structures in our setting of a real Grassmann functional integral and a real classical wave function. In sect. \ref{Massive Majorana Fermions} we finally add a mass term. We present some explicit instructive examples for the classical probability distributions that represent the propagation of massive Majorana spinors. 

In sect. \ref{Diracfermions} we turn to Dirac spinors. They are simply represented as two Majorana spinors. The electromagnetic gauge transformations rotate the two Majorana spinors into each other. The coupling to an external electromagnetic field involves therefore both types of Majorana spinors. We discuss the complex structure, the usual Weyl representation of Dirac spinors and the symmetries of the model. From the Grassmann functional integral we extract again the evolution equation for the Grassmann wave function. The associated evolution equation for the probability distribution of the classical statistical ensemble for Ising-spins describes the dynamics of an arbitrary number of electrons and positrons in an external electromagnetic field. Sect. \ref{Quantum mechanics for particle in a potential} discusses the one-particle states. The classical eight-component real wave function obeys the Dirac equation in an external electromagnetic field. The complex representation and the non-relativistic limit are straightforward. This results in the Schr\"odinger equation for a particle in a potential. All characteristic features of quantum mechanics, as interference, tunneling or discrete energy levels can therefore be derived from our evolution equation for a classical statistical ensemble of Ising-spins. This is one more clear concrete example that quantum physics can be described by classical statistics! Even though this is not the main purpose of the present paper, we comment how the discrete particle properties in quantum mechanics are related to the discrete Ising-spins, and how the continuity of the wave function simply reflects the continuity of the classical probability distribution. Particle-wave duality appears as a simple consequence of our classical statistical setting. Our conclusions are presented in sect. \ref{Conclusions}.

\section{Probability distribution and wave function}
\label{Probability distribution and wave function}
A classical statistical ensemble is specified by its states $\tau$ and a probability distribution $\{p_\tau\}$, which associates to every $\tau$ a positive probability $p_\tau\geq 0$. The distribution is normalized $\Sigma_\tau p_\tau=1$. Classical observables take in every state $\tau$ a fixed value $A_\tau$, and the mean value obeys $\kl A\kr=\Sigma_\tau A_\tau p_\tau$. 

\bigskip\noindent
{\bf 1. Generalized Ising model}

\medskip
We will consider an Ising-model type system for discrete Ising-spins $s_\gamma(\vec x)=\pm 1$, with $\vec x$ points on a suitable three-dimensional lattice and $\gamma$ denoting different ``species'' of Ising-spins, $\gamma=1\dots N_s$. For $N_s=4$ our system will be equivalent to a quantum field theory for Majorana or Weyl spinors, while for $N_s=8$ we will describe Dirac spinors. The states $\tau$ are sequences or configurations of Ising-spins, $\tau=\{s_\gamma(\vec x)\}$. Instead of Ising-spins, we will actually use occupation numbers or bits $n_\gamma(\vec x)=\big(s_\gamma(\vec x)+1\big)/2$, such that the states $\tau$ describe bit sequences of numbers $n_\gamma(\vec x)=0,1~,~\tau=\big\{n_\gamma(\vec x)\big\}$. We will consider a setting with $L^3/8$ lattice points ($L$ even) such that the number of states is $2^{(N_sL^3/8)}$. The continuum limit $L\to\infty$ is taken at the end. 

The sequences or configurations of occupation numbers $\big\{n_\gamma(\vec x)\big\}$ show already a strong analogy to the basis states of a quantum theory for an arbitrary number of fermions, formulated in the occupation number basis for positions. We will exploit this analogy in order to formulate a fundamental ``evolution law'' for the time evolution of the probability distribution $p_\tau(t)$, such that our system describes a relativistic quantum field theory for fermions.

We emphasize that the form of the evolution law is not known a priori - we do not introduce any underlying deterministic theory of ``classical trajectories'' or similar concepts. We note that no continuous time evolution for individual Ising-spins $s_\gamma(x)$ can be formulated, since $s_\gamma(x)$ only admits discrete values $\pm 1$. On the other hand, the time evolution of the probability density may well be continuous. We postulate that the basic dynamics describes the time evolution of probabilities. The corresponding dynamical law must be specified in order to define the model. The only constraint to be imposed a priori is that is respects for all $t$ the positivity and normalization of the classical probability distribution
\be\label{N11}
p_\tau(t)\geq 0~,~\sum_\tau p_\tau(t)=1.
\ee

\bigskip\noindent
{\bf 2. Classical wave function}

\medskip
A useful concept for the construction of a consistent time evolution law is the classical wave function \cite{CWP}, 
\be\label{35A}
q_\tau(t)=s_\tau(t)\sqrt{p_\tau(t)}~,~p_\tau(t)=q^2_\tau(t)~,~s_\tau(t)=\pm 1.
\ee
This is a real function which is given by the square roots of the probabilities up to signs $s_\tau(t)$. Consistent evolution laws correspond to rotations of the vector $q_\tau(t)$, 
\ba\label{35B}
q_\tau(t)=\sum_\rho R_{\tau\rho}(t,t')q_\rho(t'),\nn\\
\sum_\rho R_{\tau\rho}(t,t')R_{\sigma\rho}(t,t')=\delta_{\tau\sigma}.
\ea
The positivity of $p_\tau=q^2_\tau$ is automatic, and the normalization of the distribution remains preserved provided it was normalized for some initial time $t_{in}$, since the length of a vector or $\Sigma_\tau q^2_\tau=\Sigma_\tau p_\tau$ is preserved by rotations. We will specify a linear evolution for the wave function. In other words, we will consider rotation matrices $R_{\tau\rho}$ that are independent of the wave function.

The classical wave function specifies the probability distribution uniquely. The specification of an evolution law for the wave function therefore defines the dynamics of the classical statistical system completely. In the other direction, different choices of the sign functions $s_\tau(t)$ correspond to a choice of gauge. For all expectation values and correlations that can be computed from the probability distribution $\big\{p_\tau(t)\big\}$ the gauge choice does not matter. Natural gauge choices preserve the continuity and differentiability properties of the wave function by avoiding arbitrary discrete jumps which would be induced by arbitrary jumps in the sign functions \cite{CWP}. In contrast to the discussion of classical mechanics in a Hilbert space by Koopman and von Neumann \cite{Kop}, the classical wave function is real, such that at this step no phases appear as new degrees of freedom beyond the probability distribution. 

Quantum wave functions are usually defined in a complex Hilbert space. Indeed, many characteristic quantum features are closely associated to the ``physics of phases''. In our setting we will define a complex quantum  wave function by introducing a complex structure in the real space spanned by $\{q_\tau\}$. The $2^{(N_sL^3/8)}$ real components of the vector $\{q_\tau\}$ can then be associated to a complex vector with dimension $2^{(N_sL^3/8)-1}$. One complex structure can be closely related to the equivalence between Majorana and Weyl spinors in four dimensions \cite{CWMS}, \cite{CWMSab} \cite{CWMSa}, but more general complex structures are possible. There is no conceptual difference between the classical and quantum wave function in our setting. 

\bigskip\noindent
{\bf 3.  Grassmann wave function}

\medskip
We will specify the time evolution of the wave function $\big\{q_\tau(t)\big\}$ within a formalism based on a real Grassmann algebra. This will make the close connection to fermions most transparent \cite{CWF}. The Grassmann formulation relies on the isomorphism between states $\tau$ and the basis elements $g_\tau$ of a Grassmann algebra that can be constructed from the Grassmann variables $\psi_\gamma(x)$. Each basis element $g_\tau$ is a product of Grassmann variables
\be\label{N12}
g_\tau=\psi_{\gamma_1}(x_1)\psi_{\gamma_2}(x_2)\dots
\ee
which is ordered in some convenient way. To be specific, we define some linear ordering of the lattice points and place variables  with ``smaller $x$'' to the left, and for each $x$ place smaller $\gamma$ to the left. If a Grassmann element $g_\tau$ contains a given variable $\psi_\gamma(x)$ we put the number $n_\gamma(x)$ in the sequence $\tau$ to $0$, while we take $n_\gamma(x)=1$ if the variable $\psi_\gamma(x)$ does not appear in the product \eqref{N12}. This specifies the map between the $2^{(N_sL^3/8)}$ independent basis elements $g_\tau$ of the Grassmann algebra and the states $\tau$.

An arbitrary element $g$ of the Grassmann element can be expanded in terms of the basis elements 
\be\label{YA}
g=\sum_\tau q_\tau g_\tau.
\ee
A time dependent wave function $\big\{q_\tau(t)\big\}$ can therefore be associated to a time dependent element of the Grassmann algebra $g(t)$, provided the coefficients $q_\tau(t)$ are real and normalized according to $\sum_\tau q^2_\tau(t)=1$. We will formulate the fundamental evolution law as an evolution law for the ``Grassmann wave function'' $g(t)$. For this purpose we will formulate in the next section a Grassmann functional integral for a quantum field theory of massless Majorana or Weyl fermions. It will involve $N_sL^3/8$ Grassmann variables $\psi_\gamma(t,x)$ for every discrete time point $t$. In section \ref{Grassmann wave function} we will present a prescription how the Grassmann wave function $g(t)$ and the fundamental evolution law can be extracted from this functional integral. This will establish a map between a quantum field theory for fermions and Ising type classical statistical ensembles with dynamics specified by an appropriate evolution law. 

\section{Massless Majorana fermions}
\label{Massless Majorana fermions}

{\bf 1. Action}

\medskip
In this section we formulate the quantum field theory of a free Majorana spinor for a discrete space lattice on a torus. It will be defined by a Grassmann functional integral based on a real Grassmann algebra, and we take $N_s=4$. Due to the finite number of $(L/2)^3$ space points we have at any given $t$ a finite number $L^3/2$ of Grassmann variables $\psi_\gamma(t,\vec x)$, with four ``species'' $\gamma=1\dots 4$. (No complex conjugate Grassmann variables are defined at this stage.) For the associated classical statistical ensemble the states correspond to sequences of $L^3/2$ occupation numbers $n_\gamma(\vec x)$ that can take values $n_\gamma(\vec x)=0,1$. The total number of classical states equals $2^{L^3/2}$ and remains finite for finite $L$.

We also will take time $t$ on a finite discrete chain. The functional integral will therefore involve a finite number of Grassmann variables. It is fully regularized and all quantities are well defined. The continuum limit $L\to\infty$ of an infinite number of Grassmann variables will only be taken at the end. 

We start with the action for the regularized quantum field theory
\be\label{N1}
S=\sum^{t_f-\epsilon}_{t=t_{in}}L(t),
\ee
with Lagrangian
\be\label{A17}
L(t)=\sum_x\psi_\gamma(t,x)B_\gamma(t+\epsilon,x).
\ee
Here we define
\be\label{7A}
B_\gamma(t+\epsilon,x)=\sum_{\{v\}}Y_{\gamma\delta}\big(\{v\}\big)\psi_\delta(t+\epsilon,x_k+v_k\Delta),
\ee
with
\be\label{7B}
\sum_{\{v\}}=\prod_j\Big(\sum_{v_j=\pm 1}\Big)
\ee
and
\be\label{B17}
Y_{\gamma\delta}\big(\{v\})=\frac18
\Big[1-\sum_k(v_k+w_k\tilde I)T_k-v_1v_2v_3\tilde I\Big]_{\gamma\delta}.
\ee
Thus $B_\gamma(t+\epsilon,x)$ involves a linear combination of Grassmann variables at lattice sites which are diagonal neighbors of $x$, corresponding to corners of a cube with basis length $2\Delta$, with $x$ being at its center. The sums extend over $j,k,l=1\dots 3$ and each corner corresponds to a particular combination of the three signs $v_j=\pm 1$. We define $w_k$ by $w_1=-v_2v_3,w_2=v_1v_3$, $w_3=-v_1v_2$. Eqs. \eqref{A17}, \eqref{B17} and the following imply a summation over repeated indices $\gamma,\delta=1\dots 4$. 

We have also introduced the three real symmetric $4\times 4$ matrices $T_k$ as 
\be\label{18AA}
T_1={0,1\choose 1,0}~,~T_2={~~0,c\choose -c,0}~,~
T_3={1,~~0\choose 0,-1}~,~T^T_k=T_k,
\ee
with
\be\label{18AB}
c={~~0,1\choose -1,0}=i\tau_2,
\ee
and $1$ stands for the unit $2\times 2$ matrix. The product of these matrices yields the real antisymmetric $4\times 4$ matrix 
\be\label{18ABa}
\tilde I=-{c,0\choose~ 0,c}=T_1T_2T_3~,~\tilde I^T=-\tilde I.
\ee

The sum in eq. \eqref{N1} extends over discrete time points $t_n$, with $\int_t=\epsilon\sum_t=\epsilon\sum_n~,~t_{n+1}-t_n=\epsilon,n\in{\mathbbm Z}$, $t_{in}\leq t_n\leq t_f$. The time-continuum limit is taken as $\epsilon\to 0$ for fixed $t_{in},t_f$. Similarly, we sum in eq. \eqref{A17} over points $x$ of a lattice. For this purpose we consider for every given $t$ a cubic lattice with lattice distance $2\Delta$ and $\int_x=8\Delta^3\sum_x$. We take $\epsilon =\Delta$ and place the space-time lattice points on a hypercubic bcc lattice with distance $2\Delta$ between nearest neighbors, which we call the ``fundamental lattice''. For even $t=2n\epsilon$ the space lattice points are even, $x_k=2m'_k\Delta$, with $n,m'_k\in{\mathbbm Z}$. This will be called the even sublattice. The odd sublattice consists of the odd time points $t=(2n+1)\epsilon$ for which the space points are also odd, $x_k=(2m'_k+1)\Delta$.  The action \eqref{N1} involves indeed only Grassmann variables for points on the fundamental lattice. For even $t$ eq. \eqref{A17} involves Grassmann variables $\psi_\gamma(t,x)$ living on the even sublattice, while the combination $B_\gamma(t+\epsilon,x)$ lives on the odd sublattice. For odd $t$ the role of the sublattices is exchanged, now with $\psi$ on the odd and $B$ on the even sublattice. This construction eliminates ``lattice doublers'' - a more detailed discussion of the lattice implementation will be given elsewhere. 

The action \eqref{N1} is an element of a real Grassmann algebra - all coefficients multiplying the Grassmann variables $\psi_\gamma(t,x)$ are real. Within the Grassmann algebra the operation of transposition amounts to a total reordering of all Grassmann variables. The action \eqref{N1} is antisymmetric under this operation,
\be\label{F3}
S^T=-S.
\ee
If we define formally the Minkowski action $S_M=iS$ the latter is hermitean, $S_M=S^\dagger_M$ since $S^*_M=-S_M$. 

As an important ingredient for the probabilistic interpretation and time evolution discussed in the next two sections we observe that one can obtain $B_\gamma$ from $\psi_\gamma$ by a rotation
\ba\label{N8}
&&B_\gamma(x)=\sum_y\bar R_{\gamma\delta}(x,y)\psi_\delta(y),\nn\\
&&\sum_y\bar R_{\eta\delta}(z,y)
\bar R_{\gamma\delta}(x,y)=\delta_{\eta\gamma}\delta(z,x).
\ea
In eq. \eqref{N8} $B_\gamma$ and $\psi_\delta$ refer to the same time arguments, but we note that $x$ is not a point on the space-lattice on which the Grassmann variables $\psi_\delta(y)$ live. The quantity $B_\gamma(x)$ is a linear combination of Grassmann variables living on the corners of a cube with center at $x$. Since the number of lattice points $y$ and the number of centers of cubes $x$ is the same, we may nevertheless interpret $\bar R_{\gamma\delta}(x,y)$ as a quadratic matrix. We may formally extend the space to objects living on a cubic lattice with lattice distance $\Delta$ for each $t$. In this extended space $\bar R$ acts as a rotation matrix despite the fact that $x$ and $y$ refer to points on different sublattices of the fundamental lattice. The second equation \eqref{N8} and the following equations should be interpreted in this sense. 

In order to show that $\bar R$ is on orthogonal matrix we write it as a product
\be\label{A25}
\bar R=\tilde R_1\tilde R_2\tilde R_3,
\ee
with 
\be\label{B25}
\tilde R_k=D^+_k-T_k D^-_k.
\ee
Here the shift operators
\be\label{C25}
D^\pm_k(x,y)=\frac12\big[\delta(x,y-\Delta_k)\pm\delta
(x,y+\Delta_k)\big]
\ee
act as
\be\label{D25}
\sum_yD^\pm_k(x,y)\psi(y)=\frac12\big[\psi(x+\Delta_k)\pm\psi(x-\Delta_k)\big].
\ee
They obey
\ba\label{22A}
\sum_y D^\pm_k(x,y)D^\pm_k(y,z)&=&\\
\frac14\big[\delta(x,z-2\Delta_k)
&+&\delta(x,z+2\Delta_k)\pm 2\delta(x,y)\big],\nn
\ea
and
\ba\label{E25}
&&\sum_y D^+_k(x,y)D_k^-(y,z)=\sum_yD^-_k(x,y)D^+_k(y,z)=\nn\\
&&\quad\frac14\big[\delta(x,z-2\Delta_k)-\delta(x,z+2\Delta_k)\big].
\ea
With
\be\label{F25}
(D^\pm_k)^T=\pm D^\pm_k~,~\tilde R^T_k=D^+_k+T_kD^-_k,
\ee
one finds indeed $\tilde R^T_k\tilde R_k=1$. Observing that all shift operators $D^\pm_k,D^\pm_l$ mutually commute, one easily verifies that $B$ in eq. \eqref{B17} indeed obeys 
\be\label{20A}
B=\tilde R_1\tilde R_2\tilde R_3\psi. 
\ee

We next discuss the continuum limits in space and time for the action \eqref{N1}-\eqref{B17}. The space continuum limit $\Delta\to 0$ is characterized by
\be\label{G25}
(D^+_k\psi)(x)\to\psi(x)~,~(D^-_k\psi)(x)\to\Delta\partial_k\psi(x),
\ee
with $\partial_k=\partial/\partial x_k$ the continuum derivative. In the continuum limit one finds
\be\label{H25}
\tilde R_k\to 1-\Delta T_k\partial_k
\ee
and 
\be\label{I25}
B_\gamma(x)=\psi_\gamma(x)-\Delta\sum_k(T_k)_{\gamma\delta}
\partial_k\psi_\delta(x)+0(\Delta^2).
\ee
We further employ
\be\label{B3}
\partial_t\psi(t)=\frac1\epsilon\big[\psi(t+\epsilon)-\psi(t)\big],
\ee
such that the Grassmann property $\psi^2_\gamma(t,x)=0$ results in 
\be\label{N4}
\psi_\gamma(t,x)\partial_t\psi_\gamma(t,x)=\frac1\epsilon ~ \psi_\gamma(t,x)\psi_\gamma(t+\epsilon,x).
\ee
For $\Delta=\epsilon$ we obtain the continuum relation 
\be\label{J25}
\psi(t,x)B(t+\epsilon,x)=\epsilon(\psi\partial_t\psi-\sum_k\psi T_k\partial_k\psi).
\ee
The factor of $\epsilon$ combines with $\Sigma_t$ into $\int_t=\epsilon\Sigma_t$. Finally, the factor $8\Delta^3$ in the relation between $\int_x$ and $\sum_x$ is absorbed in the continuum limit by a rescaling of the Grassmann variables by a factor $(2\Delta)^{(-3/2)}$. 

\bigskip\noindent
{\bf 2. Lorentz symmetry}

\medskip
We next show that the action \eqref{N1} is Lorentz-symmetric in the continuum limit $\epsilon=\Delta,\Delta\to 0$. In the continuum limit we write the action as
\be\label{F1}
S=\int_{t,x}\big\{\psi_\gamma\partial_t\psi_\gamma-\psi_\gamma
(T_k)_{\gamma\delta}
\partial_k\psi_\delta\big\}.
\ee
It involves now four Grassmann functions $\psi_\gamma(t,x),\gamma=1\dots 4,x=(x_1,x_2,x_3)$. The integral extends over three dimensional space and time, with $\partial_t=\partial/\partial t$ and $\partial_k=\partial/\partial x_k$.  

The Lorentz invariance of the action \eqref{F1} is most easily established by employing the real matrices
\be\label{F7}
\gamma^0=\left(\begin{array}{rcr}
0&,&\tau_1\\-\tau_1&,&0\end{array}\right)~,~
\gamma^k=-\gamma^0 T_k,
\ee
such that
\be\label{F8}
S=-\int_{t,x}\bar\psi\gamma^\mu \partial_\mu \psi~,~\bar\psi=\psi^T\gamma^0,
\ee
where $\mu=(0,k)$ and $\partial_0=\partial_t$. We define $\bar\psi_\gamma=\psi_\delta(\gamma^0)_{\delta\gamma}$. (We consider $\psi$ here as a vector with components $\psi_\gamma$ and suppress the vector indices. The Pauli matrices are denoted by $\tau_k$.) The real $4\times 4$ Dirac matrices $\gamma^\mu$ obey the Clifford algebra 
\be\label{F9}
\{\gamma^\mu,\gamma^\nu\}=2\eta^{\mu\nu},
\ee
with signature of the metric given by $\eta_{\mu\nu}=diag(-1,1,1,1)$. This can be easily verified by using the relations
\ba\label{F10}
\{T_k,T_l\}=2\delta_{kl}~,~
\{\gamma^0,T_k\}=0.
\ea
Furthermore, one finds
\be\label{F11}
(\gamma^0)^T=-\gamma^0~,~(\gamma^k)^T=\gamma^k,
\ee
and the relations
\ba\label{F12}
[T_k,T_l]&=&2\epsilon_{klm}\tilde I T_m~,~[T_k, \tilde I]=0~,~\tilde I^2=-1,\nn\\
\gamma^0\gamma^1\gamma^2\gamma^3&=&\tilde I~,~\{\gamma^0,\tilde I\}=0~,~
\{\gamma^k,\tilde I\}=0.
\ea

Infinitesimal Lorentz-transformations can be written as
\be\label{F4}
\delta\psi_\gamma=-\frac12
\epsilon_{\mu\nu}\left(\Sigma^{\mu\nu}\right)_{\gamma\delta}\psi_\delta~,~\epsilon_{\mu\nu}=-\epsilon_{\nu\mu},
\ee
where we have omitted to indicate the associated transformations of the coordinates. The Lorentz generators $\Sigma^{\mu\nu}$ obtain from the Dirac matrices as
\be\label{F13}
\Sigma^{\mn}=-\frac14[\gamma^\mu,\gamma^\nu],
\ee
and obey
\be\label{F5}
\Sigma^{0k}=-\frac12 T_k~,~\Sigma^{kl}=-\frac12\epsilon^{klm}\tilde I T_m,
\ee
We recognize in eq. \eqref{F8} the standard Lorentz invariant action for free Majorana fermions in a Majorana representation of the Clifford algebra with real $\gamma^\mu$-matrices. 

\bigskip\noindent
{\bf 3. Functional integral}

\medskip
The functional integral is defined by the partition function
\be\label{N9}
Z=\int{\cal D}\psi\bar g_f\big[\psi(t_f)\big]e^{-S}
g_{in}\big[\psi(t_{in})\big],
\ee
with the functional measure
\be\label{N10}
\int{\cal D}\psi=\prod_{t,x}\int \big(d\psi_4(t,x)\dots d\psi_1(t,x)\big).
\ee
The boundary terms $g_{in}$ and $\bar g_f$ only depend on the Grassmann variables $\psi(t_{in})$ and $\psi(t_f)$, respectively. As we will see below, the boundary terms $\bar g_f$ and $g_{in}$ are related to each other, such that the functional integral \eqref{N9} is fully specified by the choice of $g_{in}$. 

The basic definition of $Z$ is formulated on the discrete space-time lattice with finite volume. Thus $Z$ is a well defined real number. We will show later that for a suitable normalization of $g_{in}$ one obtains $Z=1$, independently of $L$. The continuum limit $L\to\infty$ for the partition function is therefore trivial. 

We will use the Grassmann function integral \eqref{N9} in order to specify the fundamental evolution law for the probability distribution $\big\{p_\tau(t)\}$ of the classical statistical ensemble for Ising-spins. The positivity and normalization of the probabilities holds for an arbitrary choice of $g_{in}$, provided $\bar g_f$ is related to $g_{in}$ appropriately. For every given $g_{in}$ the probability distribution $\{p_\tau(t)\}$ is uniquely computable for all $t$, such that the functional integral \eqref{N9} indeed specifies the time evolution of the probability distribution. 

\section{Grassmann wave function from functional integral}
\label{Grassmann wave function}

In this section we compute for the functional integral \eqref{N9} a Grassmann wave function $g(t)$, which is an element of the Grassmann algebra constructed from the Grassmann variables $\psi_\gamma(t,x)$ at given $t$. The central idea is  to ``integrate out'' the Grassmann variables at times $t'\neq t$ \cite{14}. The expansion coefficients of $g(t)$ will specify the classical wave function $\big\{q_\tau(t)\big\}$.

\bigskip\noindent
{\bf 1. Integrating out past and future}

\medskip
The functional integral \eqref{N9} involves variables for arbitrary time points $t_n$. In order to construct a wave function $g(t)$ which only refers to a particular time $t$ we have to ``integrate out'' the information referring to other time points $t'\neq t$ \cite{CWMSa,14}. This can be done by decomposing the action \eqref{N1}
\ba\label{M1}
S&=&S_<+S_>,\nn\\
S_<&=&\sum_{t'<t}L(t')~,~S_>=\sum_{t'\geq t}L(t').
\ea
The wave function $g(t)$ obtains now by integrating out all Grassmann variables for $t'<t$
\be\label{M2}
g(t)=\int {\cal D}\psi(t'<t)e^{-S_<}g_{in}.
\ee
We observe that $g(t)$ depends only on the Grassmann variables $\psi(t)$. More precisely, it is an element of the Grassmann algebra that can be constructed from the Grassmann variables $\psi_\gamma(t,x)$. 

The conjugate wave function is defined as
\be\label{M4}
\tilde g(t)=\int {\cal D} \psi(t'>t)\bar g_f e^{-S_>}.
\ee
Again, this is an element of the Grassmann algebra constructed from $\psi(t)$. In terms of $g$ and $\tilde g$ the partition function reads
\be\label{M5}
Z=\int{\cal D}\psi(t)\tilde g(t)g(t),
\ee
where the Grassmann integration $\int {\cal D}\psi(t)$ extends now only over the Grassmann variables $\psi_\gamma(t,x)$ for a given time $t$. 

\bigskip\noindent
{\bf 2.   Classical probabilities and wave function}

\medskip
We next expand $g$ in terms of the basis elements $g_\tau$ of the Grassmann algebra generated by the variables $\psi_\gamma(t)$,
\be\label{M3}
g(t)=\sum_\tau q_\tau(t)g_\tau\big[\psi(t)\big].
\ee
We associate the real coefficients $q_\tau(t)$ with the classical wave function, such that the classical probabilities obtain as $p_\tau(t)=q^2_\tau(t)$. This requires for every $t$ the normalization $\sum_\tau q^2_\tau(t)=1$. We will show in the next section that this normalization condition is indeed obeyed, provided it holds for the initial wave function $g(t_{in})=g_{in}\big[\psi_\gamma(t_{in},x)\big]$. 

The conjugate basis elements of the Grassmann algebra $\tilde g_\tau$ are defined \cite{CWF} by the relation
\be\label{M6}
\tilde g_\tau g_\tau=\prod_x\prod_\gamma\psi_\gamma(x)
\ee
(no sum over $\tau$) and the requirement that no variable $\psi_\gamma(x)$ appears both in $\tilde g_\tau$ and $g_\tau$. They obey
\be\label{M6A}
\int{\cal D}\psi(t)\tilde g_\tau\big[\psi(t)\big]g_\rho\big[\psi(t)\big]=\delta_{\tau\rho}.
\ee
Expanding
\be\label{M7}
\tilde g(t)=\sum_\tau\tilde q_\tau(t)\tilde g_\tau\big[\psi(t)\big]
\ee
yields
\be\label{M8}
Z=\sum_\tau\tilde q_\tau(t)q_\tau(t).
\ee
We will see in the next section that for a suitable choice of $\bar g_f$ the conjugate wave function obeys for all $t$ the relation $\tilde q_\tau(t)=q_\tau(t)$. Together with the normalization condition $\sum_\tau q^2_\tau=1$ this guarantees $Z=1$.

In consequence, we can express the classical probabilities $p_\tau(t)$ directly in terms of the Grassmann functional integral 
\be\label{M9}
p_\tau(t)=\int {\cal D}\psi\bar g_f{\cal P}_\tau(t)e^{-S}g_{in},
\ee
with ${\cal P}_\tau(t)$ a projection operator
\ba\label{M10}
{\cal P}_\tau(t)g_\rho\big[\psi(t)\big]
&=&g_\tau\big[\psi(t)\big]\delta_{\tau\rho},\nn\\
\int\cD\psi(t)\tilde g_\sigma\big[\psi(t)\big]\cP_\tau(t)g_\rho\big[\psi(t)\big]
&=&\delta_{\tau\sigma}\delta_{\tau\rho}.
\ea
The formal expression of $\cP_\tau$ in terms of the Grassmann variables $\psi(t)$ and derivatives $\partial/\partial\psi(t)$ can be found in ref. \cite{3A}.

The interpretation of the Grassmann functional integral in terms of classical probabilities is based on the map $g(t)\to \big\{ p_\tau(t)\big\}$, which in turn is related to the map $\{ q_\tau\}\to \{p_\tau\}=\{ q^2_\tau\}$. The map $g(t)\leftrightarrow\big\{q_\tau(t)\big\}$ is invertible. A given state or classical statistical ensemble may be specified by the ``initial value'' at some time $t_0, g(t_0)$, or the associated wave function $\big\{q_\tau(t_0)\big\}$ or classical probability distribution $\{p_\tau(t_0)\big\}$. This is equivalent to the specification of $g_{in}=g(t_{in})$ in the functional integral.

\section{Time evolution}
\label{Timeevolution}

In this section we compute the time evolution of the wave function $\big\{q_\tau(t)\big\}$ and the associated probability distribution $\{p_\tau(t)\big\}=\big\{q^2_\tau(t)\big\}$. This will lead to a type of generalized Schr\"odinger equation for the real wave function $\big\{q_\tau(t)\big\}$, as well as an associated evolution equation for the Grassmann wave function $g(t)$. We will use the properties of this evolution equation in order to establish that the norm $\sum_\tau q^2_\tau(t)$ is conserved, such that  $\big\{q^2_\tau(t)\big\}$ can indeed be interpreted as a time dependent probability distribution. 

Due to the particular form of the action \eqref{N1}, which only involves one type of Grassmann variables (and no conjugate variables as in ref. \cite{CWF}) the discrete formulation of the functional integral is crucial. We will see that $g(t)$ jumps between two neighboring time points, while it is smooth between $g(t+2\epsilon)$ and $g(t)$. We will therefore distinguish between even and odd time points and use the definition \eqref{M3} of the wave function only for even times. (For odd times one may employ a different definition, which will guarantee the smoothness of the time evolution of $\big\{q_\tau(t)\big\}$ for both even and odd time points \cite{CWMSa}.) In the continuum limit, $\epsilon\to 0$, the time evolution of the wave function is described by a continuous rotation of the vector $\big\{q_\tau(t)\big\}$. 

As a consequence of its definition \eqref{M3} the Grassmann wave function obeys the time evolution
\be\label{P1}
g(t+\epsilon)=\int\cD\psi(t)e^{-L(t)}g(t).
\ee
This determines $\big\{q_\tau(t+\epsilon)\big\}$ in terms of $\big\{q_\tau(t)\big\}$. Thus the action \eqref{N1}-\eqref{B17} specifies the dynamics how the probability distribution $\big\{p_\tau(t)\big\}$ evolves in time. The particular dynamics of a given model is determined by the form of $B_\gamma$ in eq. \eqref{B17}.

\bigskip\noindent
{\bf   1. Unitary time evolution}

\medskip
We want to establish the relation \eqref{35B} which guarantees the preservation of the sum of probabilities $\sum_\tau p_\tau=\sum_\tau q^2_\tau=1$, with $t'\to t,~t\to t+2\epsilon$. For this purpose we employ the property \eqref{N8}. The relation between $q_\tau(t+2\epsilon)$ and $q_\tau(t)$ is computed from eq. \eqref{P1}.

Inserting the specific form \eqref{A17} for $L(t)$ the evolution equation \eqref{P1} reads
\be\label{P2}
g(t+\epsilon)=\int\cD\psi(t)\exp\big\{-\sum_x\sum_\gamma\psi_\gamma(t,x)
B_\gamma(t+\epsilon,x)\big\}g(t).
\ee
We may write $g(t)$ in a product form
\be\label{P3}
g(t)=\prod_x\prod_\gamma\big[a_\gamma(t,x)+b_\gamma(t,x)\psi_\gamma(t,x)\big],
\ee
with some fixed ordering convention of the factors assumed, e.g. smaller $\gamma$ to the left for given $x$, and some linear ordering of the lattice points, with ``lower'' points to the left. In the product form eq. \eqref{P2} yields
\ba\label{P4}
&&g(t+\epsilon)=\int\cD\psi(t)\prod_x\prod_\gamma\\
&&\Big\{\big[1-\psi_\gamma(t,x)B_\gamma(t+\epsilon,x)\big]
\big[a_\gamma(t,x)+b_\gamma(t,x)\psi_\gamma(t,x)\big]\Big\}\nn\\
&&\hspace{1.3cm}=\prod_x\prod_\gamma\Big\{b_\gamma(t,x)+\eta_\gamma a_\gamma(t,x)B_\gamma(t+\epsilon,x)\Big\}.\nn
\ea
Here we use the fact that each individual Grassmann integration $\int d\psi_\gamma(t,x)$ can be performed easily,
\be\label{P4A}
\int d\psi(1-\psi\varphi)(a+b\psi)=b-a\varphi,
\ee
and $\eta_\gamma=\pm 1$ results from the anticommuting properties of the Grassmann variables $\varphi$, with
\be\label{P5}
\eta_1=\eta_3=1~,~\eta_2=\eta_4=-1.
\ee

As a result, we can write
\be\label{P6}
g(t+\epsilon)=\sum_\tau q_\tau(t){\cal C}g_\tau \big[B_\gamma(t+\epsilon,x)\big],
\ee
where ${\cal C}g_\tau$ obtains from $g_\tau$ by the following replacements: (i) for every factor $\psi_\gamma(x)$ in $g_\tau$ one has a factor $1$ in ${\cal C}g_\tau$; (ii) for every pair $(x,\gamma)$ for which no $\psi_\gamma(x)$ is present in $g_\tau$ one inserts a factor $\eta_\gamma\psi_\gamma(x)$ in ${\cal C}g_\tau$. 
The ordering of the factors $\eta_\gamma\psi_\gamma(x)$ is the same as the ordering assumed in the product \eqref{P3}. This implies that we can indeed write the action of ${\cal C}$ on the product \eqref{P3} as
\be\label{P7}
{\cal C}g(t)=\prod_x\prod_\gamma\big(b_\gamma(t,x)+\eta_\gamma a_\gamma(t,x)\psi_\gamma (t,x)\big).
\ee

The conjugation operator ${\cal C}$ maps every basis element $g_\tau$ into its conjugate element $\tilde g_\tau$ up to a sign $\sigma_\tau=\pm 1$,
\be\label{P8}
{\cal C}g_\tau=\sigma_\tau\tilde g_\tau.
\ee
Applying ${\cal C}$ twice on the product \eqref{P3} multiplies each factor by $\eta_\gamma$. The factors $\eta_\gamma$ drop out due to the even number of minus signs, such that ${\cal C}$ is an involution, ${\cal C}^2=1$, or 
\be\label{P9}
{\cal C}^2 g_\tau=g_\tau.
\ee

The jump between $g$ and ${\cal C}g$ for neighboring time points suggests the use of eq. \eqref{M3} for the definition of the wave function $\big\{q_\tau(t)\big\}$ only for ``even time points'', namely those that obey $t_n=t_{in}+2n\epsilon,n\in{\mathbbm N}$. Repeating the procedure leading to eq. \eqref{P6} one obtains 
\be\label{58A}
g(t+2\epsilon)=\sum_\tau q_\tau(t)g_\tau\big[A_\gamma(t+2\epsilon,x)\big],
\ee
where $A_\gamma$ is a linear combination of Grassmann variables $\psi_\delta(t+2\epsilon,x)$ which live on the same sublattice of the fundamental lattice as $\psi_\delta(t,x)$. For an arbitrary linear transformation 
\be\label{58B}
B_\gamma(t+\epsilon, x)=F_{\gamma\delta}(x,y;t+\epsilon)\psi_\delta(t+\epsilon,y)
\ee
with unit Jacobian, $\det F=1$, we can write 
\ba\label{58C}
g(t+2\epsilon)&=&\int{\cal D}\psi(t+\epsilon)\nn\\
&& \exp \big\{-\psi_\gamma(t+\epsilon,x)B_\gamma(t+2\epsilon,x)\big\}g(t+\epsilon)\nn\\
&=&\int{\cal D}B(t+\epsilon)\\
&&\exp \big\{-B_\gamma(t+\epsilon,x)A_\gamma(t+2\epsilon,x)\big\}g(t+\epsilon),\nn
\ea
with 
\ba\label{58D}
A_\gamma(t+2\epsilon,x)&=&(F^{-1})^T_{\gamma\delta}(x,y;t+\epsilon)\nn\\
&& \times F_{\delta\eta}(y,z;t+2\epsilon)\psi_\eta
(t+2\epsilon,z).
\ea
Here we sum over repeated indices $\delta,\eta$ and repeated coordinates $y,z$ and write the Grassmann integration as an integration over new variables $B_\gamma(t+\epsilon,x)$. We next use the expression \eqref{P4} for $g(t+\epsilon)$ and perform the integration over $B_\gamma(t+\epsilon,x)$, 
\be\label{58E}
g(t+2\epsilon)=\prod_{x,\gamma}\eta_\gamma\big(a_\gamma(t,x)+b_\gamma(t,x)A_\gamma(t+2\epsilon,x)\big).
\ee
This replaces in $g(t)$ every factor $\psi_\gamma(t,x)$ by $A_\gamma(t+2\epsilon,x)$ such that eq. \eqref{58D} specifies the expression $A_\gamma(t+2\epsilon,x)$ appearing in eq. \eqref{58A}. 

We next employ the important property that $F$ is given by the orthogonal matrix $\bar R$ in eq. \eqref{N8}, which is independent of $t$. Since $(F^{-1})^T=F$ we end with the simple expression
\be\label{58F}
A_\gamma(t+2\epsilon,x)=\bar R^2_{\gamma\eta}(x,z)\psi_\eta(t+2\epsilon,z),
\ee
where
\be\label{58G}
\bar R^2_{\gamma\eta}(x,z)=\bar R_{\gamma\delta}(x,y)\bar R_{\delta\eta}(y,z)
\ee
defines a transformation between Grassmann variables on the same sublattice. The matrix $\bar R^2$ is, in turn, also orthogonal $(\bar R^2)^T\bar R^2=1$. We conclude that $g(t+2\epsilon)$ can be obtained from $g(t)$ by a simple rotation of the Grassmann variables.

Since $A_\gamma(x)$ is related to $\psi_\gamma(x)$ by a rotation \eqref{58F}, it is straightforward to show that $g_\tau\big[A_\gamma(x)\big]$ is also connected to $g_\tau\big[\psi_\gamma(x)\big]$ by a rotation among the basis elements
\ba\label{P16}
g_\tau\big[A_\gamma(x)\big]&=&\sum_\rho
g_\rho\big[\psi_\gamma(x)\big]R_{\rho\tau},\nn\\
\sum_\rho R_{\tau\rho}R_{\sigma\rho}&=&\delta_{\tau\sigma}.
\ea
One infers
\ba\label{P17}
g(t+2\epsilon)&=&\sum_{\tau,\rho}q_\tau(t)
g_\rho\big[\psi_\gamma(t+2\epsilon)\big]R_{\rho\tau}\nn\\
&=&\sum_\tau q_\tau(t+2\epsilon)g_\tau\big[\psi_\gamma(t+2\epsilon)\big],
\ea
with a rotated wave function 
\be\label{P18}
q_\tau(t+2\epsilon)=\sum_\rho R_{\tau\rho} q_\rho(t).
\ee
This establishes eq. \eqref{35B}. 

Rotations preserve the length of the vector $\{q_\tau\}$ such that $\sum_\tau q^2_\tau(t)$ is independent of $t$. Choosing 
$g_{in}=\sum_\tau q_\tau(t_{in})g_\tau\big[\psi(t_{in})\big]$ with $\sum_\tau q^2_\tau(t_{in})=1$ one infers $\sum_\tau q^2_\tau(t)=1$ for all $t$. Therefore $\big\{p_\tau(t)\big\}=\big\{q^2_\tau(t)\big\}$ has indeed for all $t$ the properties of a probability distribution, namely positivity of all $p_\tau$ and the normalization $\sum_\tau p_\tau=1$. In analogy to quantum mechanics we call a time evolution which preserves the norm of $\big\{q_\tau(t)\big\}$ a ``unitary time evolution''. A unitary time evolution is crucial for the probabilistic interpretation of the functional integral \eqref{N9}. 

\bigskip\noindent
{\bf  2. Evolution of conjugate wave function}

\medskip
We next want to show the relation (for $t$ even) 
\be\label{P19}
\tilde g(t-2\epsilon)=\sum_{\tau,\rho}\tilde q_\tau(t)R_{\tau\rho}\tilde g_\rho
\big[\psi_\gamma(t-2\epsilon,x)\big].
\ee
The definition \eqref{M4} of the conjugate wave function $\tilde q$ then implies
\be\label{P20}
\tilde q_\tau(t-2\epsilon)=\sum_\rho \tilde q_\rho(t)R_{\rho\tau},
\ee
such that
\be\label{P21}
\tilde q_\tau(t)=\sum_\rho R_{\tau\rho}\tilde q_\rho(t-2\epsilon).
\ee
Comparing with eq. \eqref{P18} one infers that $q_\tau(t)$ and $\tilde q_\tau(t)$ obey the same evolution equation. (By an analogous argument eq. \eqref{P21} also holds for $t$ odd.) If $q$ and $\tilde q$ are equal for some particular time $t_0$, they will remain equal for all $t$. 

If $\big\{\tilde q_\tau(t_0)\big\}$ equals $\big\{ q_\tau(t_0)\big\}$ for some time $t_0$ we can use $\tilde q_\tau(t)=q_\tau(t)$ for all $t$ and infer from eq. \eqref{M8}
\be\label{P28}
Z=\sum_\tau q^2_\tau(t).
\ee
As it should be, $Z$ remains invariant under rotations \eqref{P18} of the vector $\{q_\tau\}$ and is therefore independent of $t$. 

In order to show eq. \eqref{P19} we employ the definition \eqref{M4} which implies
\be\label{P13}
\tilde g(t-\epsilon)=\int\cD\psi(t)\tilde g(t)e^{-L(t-\epsilon)},
\ee
or
\ba\label{P22}
&&\tilde g(t-\epsilon)=\int\cD\psi(t)\tilde g(t)\exp
\big\{-\sum_x\psi_\gamma(t-\epsilon,x)B_\gamma(t,x)\big\}\nn\\
&&=\int\cD\psi(t)\tilde g(t)\exp 
\big\{-\sum_{x,y}\psi_\gamma(t-\epsilon,x)
\bar R_{\gamma\delta}(x,y)\psi_\delta(t,y)\big\}\nn\\
&&=\int{\cal D}\psi(t)\tilde g(t)\exp
\big\{-\sum_x\tilde B_\gamma(t-\epsilon,x)\psi_\gamma(t,x)\big\},
\ea
with
\ba\label{P23}
\tilde B_\gamma(t,x)&=&\sum_y\psi_\delta(t,y)\bar R_{\delta\gamma}(y,x)\nn\\
&=&\sum_y(\bar R^{-1})_{\gamma\delta}(x,y)\psi_\delta(t,y).
\ea
Performing the Grassmann integral one finds
\be\label{P24}
\tilde g(t-\epsilon)=\sum_\tau\tilde q_\tau(t)\tilde{\cal C}\tilde g_\tau
\big[\tilde B_\gamma(t-\epsilon,x)\big],
\ee
where the map $\tilde {\cal C}$ acts similarly as ${\cal C}$, with $\eta_\gamma$ replaced by $\tilde \eta_\gamma=-\eta_\gamma,
\tilde\eta_1=\tilde\eta_3=-1,\tilde\eta_2=\tilde\eta_4=1$, and $\tilde{\cal C}^2=1$. A similar step yields
\be\label{79A}
\tilde g(t-2\epsilon)=\sum_\tau\tilde q_\tau(t)\tilde g_\tau
\big[\tilde A_\gamma(t-2\epsilon,x)\big]
\ee
with 
\be\label{79B}
\tilde A_\gamma(t-2\epsilon,x)=(\bar R^2)^{-1}_{\gamma\delta}(x,y)\psi_\delta
(t-2\epsilon,y).
\ee
One concludes
\be\label{79Bb}
\tilde g_\tau\big[\tilde A_\gamma(t-2\epsilon,x)\big]
=\sum_\rho\tilde g_\rho\big[\psi_\gamma(t-2\epsilon,x)\big]R^{-1}_{\rho\tau}
\ee
and infers eq. \eqref{P19}.

\bigskip\noindent
{\bf   3. Boundary terms}

\medskip
The final point we have to settle in order to establish the normalization $Z=1$ and the expression \eqref{M9} concerns the equality of $\tilde q(t_0)$ and $q(t_0)$ for some arbitrary time $t_0$. This is achieved by a proper choice of the relation between the boundary terms $\bar g_f$ and $g_{in}$ in eq. \eqref{N9}. For this purpose we may imagine that we (formally) solve the evolution equation \eqref{P18} in order to compute $\big\{q_\tau(t_f)\big\}$ in terms of $\big\{q_\tau(t_{in}\big\}$, $g_{in}=\sum_\tau q_\tau(t_{in}) g_\tau\big[\psi(t_{in})\big]$. (We assume that $t_{in}$ and $t_f$ are even.) Using
\ba\label{P32}
g(t_f)&=&\sum_\tau q_\tau(t_f)g_\tau\big[\psi(t_f)\big], \\
\tilde g(t_f)&=&\sum_\tau\tilde q_\tau(t_f)\tilde g_\tau\big[\psi(t_f)\big]=\bar g_f,\label{P32a}
\ea
it is sufficient to choose $\bar g_f$ such that $\tilde q_\tau(t_f)=q_\tau(t_f)$. Equivalently, we may specify the wave function $\big\{q_\tau(t_0)\big\}=\big\{\tilde q_\tau(t_0)\big\}$ at some arbitrary even time $t_0$ and compute the corresponding $g_{in}$ and $\bar g_f$ by a solution of the evolution equation, using the fact that the rotation \eqref{P18} can be inverted in order to compute $q(t-\epsilon)$ form $q(t)$. 

If we would not adopt the choice $\tilde q_\tau(t_f)=q_\tau(t_f)$ the functional integral \eqref{N9} would amount to a transition amplitude, which is another useful notion in quantum field theory. We are interested, however, in a classical probabilistic setting and therefore focus on the choice of $\bar g_f$ that guarantees $Z=1$ for all $t$. In practice, neither $g_{in}$ nor $\bar g_f$ need to be computed explicitly since we only use the Grassmann functional integral for extracting the evolution equation for the Grassmann wave function $g(\tau)$ and the classical wave function $\big \{q_\tau(t)\big\}$.

\bigskip\noindent
{\bf   4. Continuous evolution equation}

\medskip
Finally, we cast the evolution law \eqref{P18} into the form of a differential time evolution equation by taking the limit $\epsilon\to 0$. This results in a generalized Schr\"odinger type equation for the real wave function $\big\{q_\tau(t)\big\}$,
\be\label{P35}
\partial_t q_\tau(t)=\sum_\rho K_{\tau\rho} q_\rho(t).
\ee
Since the evolution describes a rotation, the matrix $K$ is antisymmetric
\be\label{P36}
K_{\rho\tau}=-K_{\tau\rho}.
\ee
We identify this evolution equation with the generalized Schr\"odinger equation for a quantum wave function for the special case of a real wave function and purely imaginary and hermitean Hamiltonian $H=i\hbar K$. 

The time evolution \eqref{P35} translates directly to the probabilities (no summation over $\tau$ here)
\be\label{G14}
\partial_t p_\tau=2\sum_\rho K_{\tau\rho}s_\tau s_\rho\sqrt{p_\tau p_\rho}.
\ee
Once the signs $s_\tau(t_0)$ are fixed by some appropriate convention at a given time $t_0$, the signs $s_\tau(t)$ are computable in terms of the probabilities $p_\tau(t_0)$. This follows since for all $t$ the wave function $q_\tau(t)$ is uniquely fixed by $q_\tau(t_0)$ or $p_\tau(t_0)$, and $p_\tau(t)$ is uniquely determined by $q_\tau(t)$. In principle, it is therefore possible to formulate the time evolution law for the probabilities uniquely in terms of the probabilities $p_\tau(t_0)$. However, an expression of $\partial_tp_\tau(t)$ in terms of $p_\tau(t)$ needs to keep track of the sign functions $s_\tau(t)$. In principle, this can be done by an updating procedure. The sign $s_\tau$ can flip only at times $t$ where $p_\tau(t)=0$. If it is flipped or not at these times is decided by the requirements $p_\tau>0,\Sigma_\tau p_\tau=1$ for the following times \cite{CWMSa}. 

Instead of such an updating procedure it is much more convenient to use the wave function and the linear evolution law \eqref{P35} for the description of the classical statistical ensembles associated to the action $S$. The basic reason is the condition of unit norm of the probability distribution. This can be quite cumbersome for a general evolution equation for $\{p_\tau\}$, but it is extremely simple on the level of $\{q_\tau\}$ where only the length of a real vector has to be preserved. 

\bigskip\noindent
{\bf   5. Grassmann evolution equation}

\medskip
The matrix $K$ can be extracted from the Grassmann evolution equation
\be\label{P37}
\partial_t g={\cal K} g,
\ee
according to
\ba\label{P34}
\partial_tg(t)&=&\frac{1}{2\epsilon}\big[g(t+2\epsilon)-g(t)\big]={\cal K} g(t)=\sum_\tau q_\tau (t){\cal K} g_\tau \nn\\
&=&\sum_{\tau,\rho}q_\tau(t)g_\rho\big[\psi_\gamma(x)\big]K_{\rho\tau}=\sum_\tau\partial_t q_\tau(t)g_\tau\big[\psi_\gamma(x)\big].\nn\\
\ea
(In eq. \eqref{P34} we use a fixed basis, corresponding to the basis elements $g_\tau$ constructed from $\psi(t)$ for $g(t)$, and from $\psi(t+2\epsilon)$ for $g(t+2\epsilon)$.) For our model of free fermions the Grassmann evolution generator ${\cal K}$ is given by
\be\label{P38}
{\cal K}=\sum_x\frac{\partial}{\partial\psi_\gamma(x)}(T_k)_{\gamma\delta}
\partial_k\psi_\delta(x).
\ee
This yields the matrix element in eq. \eqref{P35}, 
\be\label{P44}
K_{\rho\tau}=\int\cD\psi\tilde g_\rho{\cal K} g_\tau.
\ee

In order to proof the relations \eqref{P37}, \eqref{P38} we 
infer from eqs. \eqref{58A}, \eqref{58F} the relation 
\be\label{89-1}
g(t+2\epsilon)-g(t)=\sum_{\tau} q_\tau(t)\big(g_\tau[\bar R^2\psi]-g_\tau[\psi]\big)
\ee
and use the continuum limit $\epsilon=\Delta\to 0$, cf. eqs. \eqref{A25}, \eqref{H25}, 
\be\label{90-1}
\bar R^{2}\psi=(1-2\epsilon\sum_k T_k\partial_k)\psi.
\ee
We then employ the identity (in linear order in $\epsilon$ and summation over $k$ implied)
\ba\label{91-1}
&&g_\tau\big[(1-2\epsilon T_k\partial_k)\psi\big]\nn\\
&&\qquad=\left(1+2\epsilon \sum_x\frac{\partial}{\partial\psi_\gamma(x)}
(T_k)_{\gamma\delta}\partial_k\psi_\delta(x)\right)g_\tau[\psi]\nn\\
&&\qquad=(1+2\epsilon{\cal K})g_\tau[\psi]
\ea
in order to establish eqs. \eqref{P37}, \eqref{P38}. 

Taking finally the space-continuum limit by rescaling $\psi_\gamma(x)$ and $\partial/\partial\psi_\gamma(x)$ such that 
\be\label{P45}
\left\{\frac{\partial}{\partial\psi_\gamma(x)}~,~
\psi_\delta(y)\right\}=
\delta_{\gamma\delta}\delta^3(x-y),
\ee
we arrive at the continuum form of the Grassmann evolution equation
\be\label{P46}
\partial_tg={\cal K} g~,~{\cal K}=\int_x
\frac{\partial}{\partial\psi_\gamma(x)}(T_k)_{\gamma\delta}
\partial_k\psi_\delta(x).
\ee
This evolution equation will be the basis for the interpretation of the time dependent wave function $\big\{q_\tau(t)\big\}$ and probability distribution $\big\{p_\tau(t)\big\}$ in terms of propagating fermionic particles.

\section{Observables}
\label{Observables}

Classical observables $A$ take a fixed value $A_\tau$ for every classical state $\tau$. In classical statistics the possible outcomes of measurements of $A$ correspond to the spectrum of possible values $A_\tau$. The expectation value of $A$ obeys 
\be\label{G15}
\kl A\kr=\sum_\tau p_\tau A_\tau.
\ee
Our description of the system will be based on these classical statistical rules. For example, we may consider the observable measuring the occupation number $N_\gamma(x)$ of the bit $\gamma$ located at $x$. The spectrum  of possible outcomes of measurements consists of values $1$ or $0$, depending if a given state $\tau=[n_\gamma(x)]$ has this particular bit occupied or empty. 

For the Grassmann basis element $g_\tau$ associated to $\tau$ one finds $N_\gamma(x)=0$ if $g_\tau$ contains a factor $\psi_\gamma(x)$, and $N_\gamma(x)=1$ otherwise. We can associate to this observable a Grassmann operator ${\cal N}_\gamma(x)$ obeying (no summation here) 
\be\label{G16}
\cN_\gamma(x)g_\tau=\big(N_\gamma(x)\big)_\tau g_\tau~,~\cN_\gamma(x)=
\frac{\partial}{\partial\psi_\gamma(x)}
\psi_\gamma(x).
\ee
Two occupation number operators ${\cal N}_{\gamma_1}(x_1)$ and ${\cal N}_{\gamma_2}(x_2)$ commute. 

In general, we may associate to each classical observable $A$ a diagonal quantum operator $\hat A$ acting on the wave function, defined by
\be\label{G16a}
(\hat A q)_\tau=A_\tau q_\tau.
\ee
This yields the quantum rule for expectation values
\be\label{G17}
\kl A\kr=\kl q\hat A q\kr=\sum_{\tau,\rho}q_\tau\hat A_{\tau\rho}q_\rho,
\ee
with $\hat A$ a diagonal operator $\hat A_{\tau\rho}=A_\tau\delta_{\tau\rho}$. In the Grassmann formulation one uses the associated Grassmann operator ${\cal A}$ obeying
\be\label{G18}
\cA g_\tau=A_\tau g_\tau,
\ee
such that 
\be\label{G19}
\kl A\kr=\int \cD\psi\tilde g\cA g.
\ee
Here $\tilde g$ is conjugate to $g$, i.e. for $g=\sum_\tau q_\tau g_\tau$ one has $\tilde g=\sum_\tau q_\tau\tilde g_\tau$.

In classical statistics the time evolution of the expectation value is induced by the time evolution of the probability distribution
\be\label{95A} 
\kl A(t)\kr=\sum_\tau p_\tau(t)A_\tau.
\ee
This corresponds to the Schr\"odinger picture in quantum mechanics
\be\label{95B}
\kl A(t)\kr=\kl q(t)\hat A q(t)\kr,
\ee
or the corresponding expression in terms of the Grassmann algebra
\be\label{95C}
\kl A(t)\kr=\int \cD\psi\tilde g(t){\cal A}g(t).
\ee

Using $\partial_t q=Kq$ \eqref{P35} and $\partial_t g={\cal K} g$ \eqref{P46} we infer for the time evolution of the expectation value the relations
\be\label{95D}
\partial_t\kl A\kr=\kl q[\hat A,K]q\kr=\int\cD\psi\tilde g[{\cal A},{\cal K}]g.
\ee
Conserved quantities are represented by Grassmann operators that commute with ${\cal K},[{\cal A},{\cal K}]=0.$

\section{Particle states}
\label{Particlestates}

Our system admits a conserved particle number, corresponding to the Grassmann operator $\cN$,
\be\label{H5}
\cN=\int_y\frac{\partial}{\partial\psi_\gamma(y)}\psi_\gamma(y)~,~[\cN,\cK]=0.
\ee
The particle number is Lorentz invariant. We can decompose an arbitrary Grassmann element into eigenstates of $\cN$
\be\label{H6}
g=\sum_mA_mg_m~,~\cN g_m=mg_m.
\ee
The time evolution does not mix sectors with different particle number $m$, such that the coefficients $A_m$ are time independent
\be\label{H7}
\partial_t g=\sum_mA_m\partial_tg_m~,~\partial_tg_m=\cK g_m.
\ee
We can restrict our discussion to eigenstates of $\cN$. The range of $m$ is $[0,B]$, with $B=N_sL^3/8=L^3/2$ the number of independent Grassmann variables. 

\bigskip\noindent
{\bf   1. Vacuum}

\medskip
Let us consider some static vacuum state $g_0$ with a fixed particle number $m_0$,
\be\label{H8}
\cK g_0=0~,~\cN g_0=m_0g_0~,~\int \cD\psi \tilde g_0 g_0=1.
\ee
An example for a possible vacuum state is the totally empty state $g_0=|0\kr$, with
\ba\label{47Aa}
|0\kr=\prod_\alpha\psi_\alpha&=&\prod_x\prod_\gamma\psi_\gamma(x)
=\prod_x(\psi_1\psi_2\psi_3\psi_4),\nn\\
\cN|0\kr&=&0.
\ea
It obeys
\be\label{104Aa}
\int\cD\psi|0\kr=1~,~|\tilde 0\kr=1.
\ee
For $m_0\neq 0$ we shift the particle number by an additive ``renormalization'' $n=m-m_0$, such that the vacuum corresponds to $n=0$, and $g=A_n g_n$. An eigenstate of ${\cal N}$ with eigenvalue $m=m_0+n$ is called a $n$-particle state if $n$ is positive, and a $n$-hole state for negative $n$. 

\bigskip\noindent
{\bf   2. One-particle and one-hole states}

\medskip
We next define creation and and annihilation operators $a^\dagger_\gamma(x),~a_\gamma(x)$ as
\be\label{47A}
a^\dagger_\gamma(x)g=\frac{\partial}{\partial\psi_\gamma(x)}g~,~a_\gamma(x)g=
\psi_\gamma(x)g.
\ee
They obey the standard (anti-)commutation relations
\ba\label{H6a}
\big \{a^\dagger_\gamma(x),~a_\epsilon(y)\big\}&=&\delta_{\gamma\epsilon}\delta(x-y)~,~
\cN=\int_xa^\dagger_\gamma(x)a_\gamma(x),\nn\\
~[a^\dagger_\gamma(x),\cN]&=&-a^\dagger_\gamma(x)~,~[a_\gamma(x),\cN]=a_\gamma(x).
\ea
Acting with the creation operator on the vacuum produces one-particle states
\ba\label{H7a}
&&g_1(t)=\int_xq_\gamma(t,x)a^\dagger_\gamma(x)g_0=\cG_1g_0,\nn\\
&&(\cN-m_0)g_1=g_1.
\ea
If needed, we may multiply ${\cal G}_1$ with an appropriate normalization factor such that the wave function obeys
\be\label{57A}
\int_x\sum_\gamma q^2_\gamma(x)=1,
\ee
and $g_1$ has a standard normalization. 

Similarly, a one-hole state with $n=-1$ obtains by employing the annihilation operator
\ba\label{H8a}
g_{-1}(t)&=&\int_x\hat q_\gamma(t,x)a_\gamma(x)g_0\nn\\
&=&-\int_x\bar q_\gamma(t,x)(\gamma^0)_{\gamma\delta}a_\delta(x)g_0=
\cG_{-1}g_0,\nn\\
&&~~(\cN-m_0)g_{-1}=-g_{-1}.
\ea
No one-hole states exist for the vacuum \eqref{47Aa}, but this issue is different if $m_0\neq 0$, as for example for $g_0=1$ where $m_0=B$.

If we transform the one-particle wave function $q_\gamma(t,x)$ infinitesimally according to 
\ba\label{H9}
\delta q_\gamma=-\frac12\epsilon_{\mn}\left (\Sigma^{\mn}\right)_{\gamma\delta}q_\delta
\ea
the operator $\cG_1$ is Lorentz invariant. (We omit here the part resulting from the change of coordinates.) Similarly, $\hat q_\gamma$ transforms as $q_\gamma$ and the corresponding infinitesimal transformation results in an invariant ${\cal G}_{-1}$. If the vacuum $g_0$ is Lorentz invariant, the Lorentz transformed one-particle or one-hole wave functions will obey the same evolution equations as the original wave functions.

The time evolution of the one particle wave function $q_\gamma$ is given by
\be\label{H11} 
\partial_t g_1=\cK g_1=\int_x(\partial_tq\frac{\partial}{\partial\psi})g_0=\int_x 
q\left[\cK,\frac{\partial}{\partial\psi}\right]g_0.
\ee
and similar for the hole. With
\be\label{54A}
\left[\cK,\frac{\partial}{\partial\psi_\gamma(x)}\right]=-
\partial_k\frac{\partial}{\partial\psi_\epsilon(x)}
(T_k)_{\epsilon\gamma}
\ee
and
\ba\label{H12}
\big[\cK,\psi_\gamma(x)\big]=-\partial_k\psi_\epsilon(x)(T_k)_{\epsilon\gamma},
\ea
one obtains Dirac equations for real wave functions (with $\partial_0=\partial_t)$
\be\label{H13}
\gamma^\mu \partial_\mu q=0~,~\gamma^\mu \partial_\mu \hat q=0.
\ee
We emphasize that these equations follow for arbitrary static states $g_0$ which obey ${\cal K}g_0=0$.

\bigskip\noindent
{\bf  3. Weyl and Majorana spinors}

\medskip
Let us consider $g_0=|0\kr$ where $a_\gamma(x)|0\kr=0$ implies that no hole states exist. There are then only particle states and the propagating degrees of freedom correspond to Majorana fermions. In four dimensions Majorana spinors are equivalent to Weyl spinors \cite{CWMS}. Indeed, we may introduce a complex structure by defining a two-component complex spinor 
\be\label{H14}
\varphi(x)=\left(\begin{array}{c}
\varphi_1(x)\\\varphi_2(x)
\end{array}\right)~,~
\varphi_1=q_1+iq_2~,~\varphi_2=q_3+iq_4.
\ee
The matrices $T_k=\gamma^0\gamma^k$ are compatible with this complex structure. They are translated to the complex Pauli matrices
\be\label{H15}
T_k=\gamma^0\gamma^k\to \tau_k,
\ee
while the operation $\varphi\to i\varphi$ corresponds to the matrix multiplication $q\to \tilde I q$. (Cf. ref. \cite{CWMSa} for more details of the map from Majorana to Weyl spinors.) The Dirac equation reads in the complex basis
\be\label{H16}
\partial_t\varphi=\tau_k\partial_k\varphi~,~i\partial_t\varphi=-\tau_k{\cal P}_k\varphi~,~{\cal P}_k\varphi=-i\partial_k\varphi,
\ee
and the Lorentz generators are given by
\be\label{H17}
\Sigma^{kl}\to-\frac i2\epsilon^{klm}\tau_m~,~
\Sigma^{0k}\to-\frac12\tau_k.
\ee

In contrast, the multiplications with $\gamma^0$ cannot be represented by a multiplication of $\varphi$ with a complex $2\times 2$ matrix, since 
\be\label{18}
q\to \gamma^0 q~~\widehat{=}~~\varphi\to -\tau_2\varphi^*.
\ee
If we express $\varphi^*$ in terms of the two-component complex vector
\be\label{H19}
\chi=E\varphi^*=-i\tau_2\varphi^*=\left(\begin{array}{r}
-q_3+iq_4\\q_1-iq_2\end{array}\right),
\ee
the transformation $q\to\gamma^0 q$ corresponds to 
\be\label{H20}
\varphi\to-i\chi~,~\chi\to -i\varphi.
\ee

We may now introduce the complex four component vector
\be\label{H21}
\Psi_M=\left(\begin{array}{c}
\varphi\\\chi\end{array}\right),
\ee
for which all matrix multiplications $q\to \gamma^\mu q$ can be represented by multiplication with complex matrices of the Clifford algebra,
\ba\label{H22}
&&\gamma^0=\left(\begin{array}{cc}0,&-i\\-i,&0\end{array}\right)~,~
\gamma^k=
\left(\begin{array}{cc}0,&-i\tau_k\\i\tau_k,&0\end{array}\right),\\
&&\Sigma^{0k}=-\frac12\left(\begin{array}{cc}
\tau_k,&0\\0,&-\tau_k
\end{array}\right)~,~
\Sigma^{kl}=-\frac i2\epsilon^{klm}
\left(\begin{array}{cc}
\tau_m,&0\\0,&\tau_m
\end{array}\right).\nn
\ea
We can also define the matrix
\be\label{H23}
\bar\gamma=-i\gamma^0\gamma^1\gamma^2\gamma^3=
\left(\begin{array}{cc}1,&0\\0,&-1\end{array}\right),
\ee
for which $\varphi$ and $\chi$ are eigenvectors with eigenvalues $\pm 1$. (Often $\bar\gamma$ is denoted as $\gamma^5$.) The transformation $q\to\tilde I q$ acts on $\psi_M$ as $\psi_M\to i\bar\gamma \psi_M$. Thus the representation \eqref{H23}, $\bar\gamma=-i\tilde I$, is consistent with the complex structure. Since $\chi$ is not independent of $\varphi$ the spinor $\Psi_M$ obeys the Majorana constraint \cite{CWMS}, \cite{CWMSab}.
\be\label{H24}
B^{-1}\Psi_M^*=\Psi_M~,~B=B^{-1}=-\gamma^2=
\left(\begin{array}{cccc}
0&0&0&1\\0&0&-1&0\\0&-1&0&0\\1&0&0&0
\end{array}\right),
\ee
where $B\gamma^\mu B^{-1}=(\gamma^\mu)^*$. 

The real Dirac matrices $\gamma^\mu_{(M)}$ in the ``Majorana representation'' \eqref{F7} and the Dirac matrices $\gamma^\mu_{(W)}$ in the ``Weyl representation'' \eqref{H22} are related by a similarity transformation,
\be\label{67A}
q=\frac{1}{\sqrt{2}}A\Psi_M~,~\gamma^\mu_{(W)}=A^{-1}\gamma^\mu_{(M)}A,
\ee
with a unitary matrix $A$
\be\label{67B}
A=\frac{1}{\sqrt{2}}
\left(\begin{array}{cccc}
1,&0,&,0,&1\\-i,&0,&0,&i\\0,&1,&-1,&0\\0,&-i,&-i,&0
\end{array}\right)~,~A^\dagger A=1.
\ee

\bigskip\noindent
{\bf   4. Weyl particles}

\medskip
The propagating degrees of freedom correspond to the solutions of the evolution equation. They are most simply discussed in terms of the complex equation \eqref{H16}. We can perform a Fourier transform
\be\label{H26}
\varphi(t,x)=\int_p\tilde \varphi(t,p)e^{ipx}=\int\frac{d^3p}{(2\pi)^3}\tilde\varphi
(t,p)e^{ipx},
\ee
such that the evolution equation becomes diagonal in momentum space,
\be\label{H27}
i\partial_t\tilde\varphi(t,p)=-p_k\tau_k\tilde\varphi(t,p).
\ee
The general solution of eq. \eqref{H16} obeys
\be\label{H28}
\tilde\varphi(t,p)=\exp(ip_k\tau_kt)\tilde \varphi(p),
\ee
with arbitrary complex two-component vectors $\tilde\varphi(p)$. This is the standard time evolution for a propagating Weyl particle.

\bigskip\noindent
{\bf   5. Multi-fermion states}

\medskip
States with arbitrary $n$ describe systems of $n$ fermions. This is not surprising in view of our translation of the classical statistical ensemble to a Grassmann functional integral. For $g_0=|0\kr$ there are only $n$-particle states, and no hole states. They can be constructed by applying $n$ creation operators $a^\dagger_\gamma(x)$ on the vacuum. For example, the two-fermion state obeys
\be\label{H33}
g_2(t)=\frac{1}{\sqrt{2}}\int_{x,y}q_{\gamma\epsilon}(t,x,y)a^\dagger_\gamma(x)
a^\dagger_\epsilon(y)g_0.
\ee
Due to the anticommutation relation
\be\label{H34}
\big\{a^\dagger_\gamma(x)~,~a^\dagger_\epsilon(y)\big\}=
\left\{\frac{\partial}{\partial\psi_\gamma(x)}~,~
\frac{\partial}{\partial\psi_\epsilon(y)}\right\}=0
\ee
the two-particle wave function is antisymmetric, as appropriate for fermions
\be\label{H35}
q_{\gamma\epsilon}(t,x,y)=-q_{\epsilon\gamma}(t,y,x).
\ee
The Grassmann elements $g_2$ describe states with two Weyl spinors. A particular class of states can be constructed as products of appropriately normalized one particle states $q^{(1)},q^{(2)}$
\be\label{H36}
q_{\gamma\epsilon}(x,y)=q^{(1)}_\gamma(x)q^{(2)}_\epsilon(y)-q^{(1)}_\epsilon(y)
q^{(2)}_\gamma(x).
\ee
General two particle states are superpositions of such states.

\bigskip\noindent
{\bf 6. Symmetries}

\medskip
Besides Lorentz symmetry the action \eqref{F1} is also invariant under global $SO(2)$ rotations
\ba\label{71A}
\psi'_1&=&\cos\alpha~\psi_1-\sin\alpha~\psi_2~,~\psi'_2
=\sin\alpha~\psi_1+\cos\alpha~\psi_2,\nn\\
\psi'_3&=&\cos\alpha~\psi_3-\sin\alpha~\psi_4~,~\psi'_4
=\sin\alpha~\psi_3+\cos\alpha~\psi_4.\nn\\
\ea
This is easily seen from the infinitesimal transformation
\be\label{71B}
\delta\psi_\gamma=\alpha\tilde I_{\gamma\delta}\psi_\delta
\ee
and the relations $\tilde I^2=-1,~\tilde I^T=-\tilde I,~[\tilde I,T_k]=0$. These rotations carry over to the one-particle and one-hole wave functions $q_\gamma$ and $\hat q_\gamma$. 

Correspondingly, the complex two component wave functions $\varphi$ and $\chi$ transform as 
\be\label{71C}
\varphi'=e^{i\alpha}\varphi~,~\chi'=e^{-i\alpha}\chi.
\ee
The $SO(2)$ rotations are now realized as $U(1)$ phase rotations. If we define for a general complex field $\eta$ the charge $\bar Q$ by the transformation 
\be\label{71D}
\eta'=e^{i\alpha \bar Q}\eta
\ee
we infer that $\varphi$ carries charge $\bar Q=1$, while $\chi$ has opposite charge $-1$. If $\varphi$ describes degrees of freedom of an electron, $\chi$ describes the corresponding ones for a positron. We note that charge eigenstates exist only in connection with a complex structure. The real wave function $q_\gamma$ can be encoded both in $\varphi$ and $\chi$ and may therefore describe degrees of freedom with opposite charge. 

The action \eqref{F1} is further invariant with respect to discrete symmetries. Among them a parity type reflection maps
\be\label{85Aa}
\psi(x)\to\bar{\cal P}\psi(x),
\ee
with
\ba\label{85Ba}
\big (\bar{\cal P}\psi(x)\big)_\gamma&=&(\gamma^0)_{\gamma\delta}\psi_\delta(-x),
\ea
and we note $\bar{\cal P}^2=-1$. We may associate $\bar{\cal P}$ with a CP-transformation for Weyl spinors. The action \eqref{F1} is further invariant under the discrete transformation $\psi\to\tilde I\psi$. It is therefore also invariant under a parity type transformation where $\bar{\cal P}$ is replaced by $\tilde I\bar {\cal P}$.

\section{Complex structure}
\label{Complex structure}

We have formulated the classical statistical description of a quantum field theory for Majorana spinors in terms of a real Grassmann algebra. All quantities in the functional integral \eqref{N9} and the action \eqref{F1} or \eqref{N1}-\eqref{B17} are real. The introduction of complex Weyl spinors in eq. \eqref{H14} or \eqref{H19} reveals the presence of a complex structure in this real formulation. Such complex structures are the basis for the importance of phases in quantum mechanics. The Schr\"odinger equation for a one-particle state can be formulated in terms of a complex wave function, and this generalizes to multi-particle states. 

\bigskip\noindent
{\bf   1. Complex Grassmann variables}

\medskip
Complex Grassmann variables may be introduced in analogy to eq. \eqref{H14} 
\be\label{71LA}
\zeta_1=\frac{1}{\sqrt{2}}(\psi_1+i\psi_2)~,~
\zeta_2=\frac{1}{\sqrt{2}}(\psi_3+i\psi_4).
\ee
Together with the complex conjugate Grassmann variables
\be\label{158A}
\zeta^*_1=\frac{1}{\sqrt{2}}(\psi_1-i\psi_2)~,~\zeta^*_2=\frac{1}{\sqrt{2}}
(\psi_3-i\psi_4)
\ee
we have for every $x$ and $t$ four independent Grassmann variables $\zeta_1,\zeta^*_1,\zeta_2,\zeta^*_2$, which replace $\psi_1,\psi_2,\psi_3,\psi_4$. 

A general element of a complex Grassmann algebra can be expanded as
\ba\label{71M}
g_c&=&\sum_{k,l}c_{\alpha_1\dots \alpha_k,\bar\alpha_1\dots\bar\alpha_l}
(x_1\dots x_k,\bar x_1\dots \bar x_l)\nn\\
&&\zeta_{\alpha_1}(x_1)\dots\zeta_{\alpha_k}(x_k)\zeta^*_{\bar\alpha_1}(\bar x_1)\dots
\zeta^*_{\bar\alpha_l}(\bar x_l),
\ea
with complex coefficients $c$. For a real Grassmann algebra the coefficients $c$ are restricted by $g^*=g$. From a general complex $g_c$ we can obtain an element of a real Grassmann algebra as
\be\label{87A}
g=\frac12(g_c+g^*_c).
\ee
The action \eqref{F1} is an element of a real Grassmann algebra and reads in terms of $\zeta$ 
\ba\label{168D}
S&=&\int_{t,x}\big\{\zeta^\dagger(\partial_t-\tau_k\partial_k)\zeta+\zeta^T
(\partial_t-\tau^*_k\partial_k)\zeta^*\big\}\nn\\
&=&2\int_{t,x}\zeta^\dagger(\partial_t-\tau_k\partial_k)\zeta.
\ea
Up to the factor $2$, which may be removed by a rescaling of $\zeta$, this is the action for a free Weyl spinor.

On every factor $\zeta$ the action of a matrix multiplication of $\psi$ by $\tilde I$ amounts to a multiplication with $i$. More precisely, we can interpret eq. \eqref{71LA} as a map $\psi\to\zeta[\psi]$ with the property $\zeta[\tilde I\psi]=i\zeta[\psi]$. For the complex conjugate one has $\zeta^*[\tilde I\psi]=-i\zeta^*[\psi]$. For the infinitesimal transformation \eqref{71B} one concludes 
\be\label{71N}
\delta\big [\zeta_{\alpha_1}(x_1)\dots \zeta^*_{\bar\alpha_l}(\bar x_l)\big ]=
i\alpha\bar Q\zeta_{\alpha_1}(x_1)\dots\zeta^*_{\bar\alpha_l}(\bar x_l),
\ee
where the charge $\bar Q$ counts the number of factors $\zeta$ minus the number of factors $\zeta^*$ for a given term in the expansion \eqref{71M}. In other words, products with $\bar Q_+$ factors $\zeta$ and $\bar Q_-$ factors $\zeta^*$ are charge eigenstates with $\bar Q=\bar Q_+-\bar Q_-$. States with a given $\bar Q$ are degenerate since many different choices of $\bar Q_+,\bar Q_-$ lead to the same $\bar Q$. For a real Grassmann algebra we use eq. \eqref{87A}. Every term in $g_c$ with $\bar Q\neq 0$ is accompanied by a term with opposite charge $-\bar{Q}$ in $g^*_c$. Thus the expansion \eqref{71M} of eq.  \eqref{87A} involves ``charge eigenspaces'' with pairs of opposite axial charge. 

It will often be convenient to use an extension to a complex Grassmann algebra, where the coefficients $c$ in eq. \eqref{71M} are arbitrary and charge eigenstates belong to the Grassmann algebra also for $\bar Q\neq 0$. This can be mapped at the end to a real Grassmann algebra by eq. \eqref{87A}. The action of $\bar Q_\pm$ on $g_c$ is represented as 
\be\label{71O}
\bar Q_+=\int_y\zeta_\alpha(y)\frac{\partial}{\partial\zeta_\alpha(y)}~,~
\bar Q_-=\int_y\zeta^*_\alpha(y)\frac{\partial}{\partial\zeta^*_\alpha(y)}.
\ee
The Grassmann derivatives $\partial/\partial \zeta_\alpha$ obey the standard anticommutation relations
\ba\label{104A}
\left\{\frac{\partial}{\partial\zeta_\alpha(x)},\zeta_\beta(y)\right\}&=&
\delta_{\alpha\beta}\delta(x-y)~,~
\left\{\frac{\partial}{\partial\zeta_\alpha(x)},\zeta^*_\beta(y)\right\}=0.\nn\\
\ea

In the complex basis we can write the time evolution equation in the form
\be\label{71Q}
i\partial_t g_c={\cal H}g_c
\ee
with 
\be\label{71R}
{\cal H}=i\int_x\left[\frac{\partial}{\partial\zeta_\alpha}
(\tau_k\partial_k\zeta)_\alpha+\frac{\partial}{\partial\zeta^*_\alpha}
(\tau^*_k\partial_k\zeta^*)_\alpha\right].
\ee
Both $\bar Q_+$ and $\bar Q_-$ commute with ${\cal H}$. The operators creating one hole or one particle states read
\ba\label{93A}
\hat q_\gamma\psi_\gamma&=&\frac{1}{\sqrt{2}}
(\hat\varphi^*_\alpha\zeta_\alpha+\hat\varphi_\alpha\zeta^*_\alpha),\nn\\
q_\gamma\frac{\partial}{\partial\psi_\gamma}&=&\frac{1}{\sqrt{2}}
\left(\varphi_\alpha\frac{\partial}{\partial\zeta_\alpha}+\varphi^*_\alpha
\frac{\partial}{\partial\zeta^*_\alpha}\right).
\ea

The infinitesimal Lorentz transformations of the complex two-component spinors $\zeta,\delta\zeta=-\frac12\epsilon_{mn}\Sigma^{mn}\zeta$, are represented by the complex $2\times 2$  matrices $\Sigma^{mn}$ given by eq. \eqref{H17}. We also observe that  $\tilde\zeta=E\zeta=-i\tau_2\zeta$ transforms as $\delta\tilde\zeta=\frac12\epsilon_{mn}(\Sigma^{mn})^T\tilde\zeta$ such that
\be\label{93B}
\frac12\tilde\zeta^T\zeta=\zeta_1\zeta_2=\frac12\big[\psi_1\psi_3-\psi_2\psi_4+i
(\psi_1\psi_4+\psi_2\psi_3)\big]
\ee
is a Lorentz scalar. The same holds for $\zeta^*_1\zeta^*_2$ such that the bilinears $\psi_1\psi_3-\psi_2\psi_4$ and $\psi_1\psi_4+\psi_2\psi_3$ are separately Lorentz scalars. Also the product
\be\label{93C}
\zeta_1\zeta_2\zeta^*_1\zeta^*_2=\psi_1\psi_2\psi_3\psi_4=
\frac{1}{24}\epsilon^{\alpha\beta\gamma\delta}
\psi_\alpha\psi_\beta\psi_\gamma\psi_\delta
\ee
is a Lorentz scalar. The functional measure obeys
\ba\label{93D}
&&\int d \psi_4 d\psi_3 d\psi_2 d\psi_1=\prod_\gamma d\psi_\gamma\nn\\
&&=\int d\zeta_2 d\zeta_1\int d\zeta^*_2 d\zeta^*_1=
\int d \zeta d\zeta^*.
\ea

\bigskip\noindent
{\bf   2. General complex structure}

\medskip
A real Grassmann algebra will, in general, admit different possible complex structures. A discussion of phases in quantum mechanics needs a specific choice of the complex structure. For example, our discussion of Dirac-spinors in sect. \ref{Diracfermions} will employ a complex structure different from eq. \eqref{71LA}. We therefore briefly discuss the general properties of complex structures. 

In a real even-dimensional vector space a complex structure is given by the existence of an involution $K$, together with a map $I$, obeying
\be\label{186A}
K^2=1~,~I^2=-1~,~\{K,I\}=0.
\ee
The matrix $K$ has eigenvalues $\pm 1$ and we may denote eigenstates with positive eigenvalues by $v_R$ and those with negative ones by $v_I,Kv_R=v_R, Kv_I=-v_I$. The matrix $I$ is a map  between $v_R$ and $v_I$, implying that the number of independent $v_R$ and $v_I$ are equal. We can choose the $v_I$ such that $Iv_R=v_I,Iv_I=-v_R$ and use these properties for defining a map from the real vectors $v$ to complex vectors $c=v_R+iv_I=c(v)$, with the properties $c(Kv)=\big[c(v)\big]^*$, $c(Iv)=ic(v)$. A linear operator or observable $A$ can be represented by multiplication with a complex matrix if $[A,I]=0$.

As an example, we consider the complex structure underlying eq. \eqref{71LA}. The maps \eqref{H14} or \eqref{71LA} act on $v=\{q_\gamma\}$ or $v=\{\psi_\gamma\}$, and the matrices $K,I$ are given by 
\be\label{186B}
K=\tilde K=\left(\begin{array}{rr}
\tau_3,&0\\0,&\tau_3
\end{array}\right)~,~I=\tilde I.
\ee
Indeed, the map $\zeta[\psi]$ given by eq. \eqref{71LA} obeys
\be\label{186C}
\zeta[\tilde K\psi]=\zeta^*[\psi]~,~\zeta[\tilde I\psi]=i\zeta[\psi].
\ee

A transformation $\psi\to A\psi$ is compatible with this complex structure if $A$ obeys
\be\label{186D}
[A,\tilde I]=0.
\ee
In this case $A$ can be represented by complex matrix multiplication acting on $\zeta$,
\be\label{186E}
\zeta[A\psi]=\tilde A\zeta[\psi].
\ee
The matrices $T_k$ commute with $\tilde I$ and are therefore compatible with the complex structure
\be\label{186F}
\zeta[T_k\psi]=\tau_k\zeta[\psi].
\ee
Also $\tilde I$ and $\Sigma^{\mn}$ (cf. eq. \eqref{F5}) are compatible with the complex structure, where the action of $\Sigma^{\mn}$ on $\zeta$ is given by eq. \eqref{H17}. The matrices $A$ obeying eq. \eqref{186D} form a group, since the product of two matrices $A_1A_2$ again commutes with $\tilde I$. This product is represented by complex matrix multiplication of $\tilde A_1$ and $\tilde A_2$, $\zeta[A_1A_2\psi]=\tilde A_1\tilde A_2\zeta[\psi]$. In contrast, the matrices $\gamma^\mu$ anticommute with $\tilde I$, cf. eq. \eqref{F12}, and are therefore not compatible with the complex structure \eqref{186B}. One finds
\be\label{186G}
\zeta[\gamma^0\psi]=-\tau_2\zeta^*.
\ee
The complex structure \eqref{186A}, \eqref{186B} for the Grassmann variables can be extended to a complex structure for the real Grassmann algebra  by defining a suitable map $g\to g_c$.

\section{Massive Majorana Fermions}
\label{Massive Majorana Fermions}

The action \eqref{F1} describes massless Majorana or Weyl spinors. For a massive Majorana spinor one adds a mass term
\ba\label{D1a}
S_m&=&\int_{t,x}\bar\psi(m\tilde I-\tilde m)\psi\nn\\
&=&\int_{t,x}\psi_\gamma\big[m(\gamma^0\tilde I)_{\gamma\delta}-\tilde m(\gamma^0)_{\gamma\delta}\big]\psi_\delta.
\ea
In the presence of the mass term \eqref{D1a} the action remains Lorentz-invariant, anti-hermitean and real (for real $m,\tilde m$), and the Minkowski action $S_M=iS$ is hermitean. For discrete time steps the first factor $\psi$ is taken at $t$, and the second at $t+\epsilon$. The $U(1)$-rotations \eqref{71A} do not leave $S_m$ invariant. Indeed, the infinitesimal transformation \eqref{71B} results in
\be\label{D3}
\tilde M\to \tilde M+\delta\tilde M~,~\delta\tilde M=-\alpha[\tilde I,\tilde M]=
2\alpha\tilde M\tilde I,
\ee
where
\be\label{D2}
\tilde M_{\gamma\delta}=m(\gamma^0\tilde I)_{\gamma\delta}-\tilde m(\gamma^0)_{\gamma\delta}.
\ee
The terms $\sim m$ and $\tilde m$ are rotated into each other,
\be\label{D4}
\delta m=-2\alpha\tilde m~,~\delta\tilde m=2\alpha m.
\ee
We may use this transformation in order to choose a convention where $\tilde m=0$. With respect to the parity transformation $\bar{\cal P}$ \eqref{85Ba} the term $\sim \tilde m$ is invariant, while $m$ changes sign. On the other hand, $S_m$ is odd with respect to the discrete transformation $\psi\to\tilde I\psi$. Thus, with respect to the parity transformation $\tilde I\bar{\cal P}$ one finds that $m$ is invariant while $\tilde m$ changes sign.

\bigskip\noindent
{\bf   1. Evolution equation}

\medskip
For massive Majorana spinors the Grassmann time evolution \eqref{P46} obeys
\be\label{D5}
{\cal K}=\int_x\frac{\partial}{\partial\psi_\gamma(x)}
\big\{(T_k)_{\gamma\delta}\partial_k-\tilde M_{\gamma\delta}\big\}\psi_\delta(x)={\cal K}_0+{\cal K}_m,
\ee
with
\be\label{D6}
{\cal K}_m=-\int_x\frac{\partial}{\partial\psi_\gamma(x)}\tilde M_{\gamma\delta}
\psi_\delta(x).
\ee
The particle number ${\cal N}$ remains conserved, while the charge $\bar Q$ is no longer a conserved quantity in the presence of a mass term. 

We may consider general operators of the type
\be\label{H1}
\cB_{\epsilon\eta}(y)=\frac{\partial}{\partial\psi_\epsilon(y)}~B_{\epsilon\eta}(y)
\psi_\eta(y),
\ee
with $B_{\epsilon\eta}$ depending on $y$ and derivatives with respect to $y$. They obey the commutation relation
\ba\label{H2}
&&\big[\cB_{\epsilon\eta}(y),\cK_0\big]=\partial_k
\left\{\frac{\partial}{\partial\psi_\gamma}(T_k)_{\gamma\epsilon}B_{\epsilon\eta}
\psi_\eta\right\}\\
&&\quad +\frac{\partial}{\partial\psi_\gamma}\big\{B_{\gamma\epsilon}
(T_k)_{\epsilon\eta}\partial_k\psi_\eta-(T_k)_{\gamma\epsilon}\partial_k(B_{\epsilon\eta}\psi_\eta)\big\},\nn
\ea
where all quantities on the r.h.s. depend on $y_k$ and $\partial_k=\partial/\partial y_{k}$. For the mass contribution one finds
\be\label{D7}
[{\cal B}_{\epsilon\eta}(y),{\cal K}_m]=\frac{\partial}{\partial\psi_\gamma(y)}
\big[\tilde M,B(y)\big]_{\gamma\delta}\psi_\delta(y).
\ee

The momentum operator \eqref{H16} reads in the real Grassmann algebra
\be\label{H4}
\cP_k=-\int_y\frac{\partial}{\partial\psi}\tilde I\partial_k\psi.
\ee
For $\tilde M\neq 0$ it is no longer conserved, while we may define conserved quantities involving an even number of derivatives
\be\label{D7A}
{\cal P}^2_k=-\int_y\frac{\partial}{\partial\psi_\gamma}\partial^2_k\psi_\gamma.
\ee

For any static vacuum state $g_0,{\cal K}g_0=0$, the evolution of the one-particle state \eqref{H7a} follows eq. \eqref{H11}
\be\label{D8}
\int_x\partial_t q_\gamma(x)\frac{\partial}{\partial\psi_\gamma(x)}g_0=\int_x q_\gamma(x)
\left[{\cal K},\frac{\partial}{\partial\psi_\gamma(x)}\right]g_0.
\ee
With 
\be\label{D9}
\left[\cK_m,\frac{\partial}{\partial\psi_\gamma(x)}\right]=-
\frac{\partial}{\partial\psi_\delta(x)}\tilde M_{\delta\gamma}=\tilde M_{\gamma\delta}
\frac{\partial}{\partial\psi_\delta(x)}
\ee
one finds the evolution equation
\be\label{D10a}
\partial_tq=(T_k\partial_k-\tilde M)q~,~
\cD q=(\gamma^\mu\partial_\mu-M)q=0,
\ee
where
\be\label{D11}
M=-\gamma^0\tilde M=m\tilde I-\tilde m.
\ee
For $\tilde m=0$ the relation $\{\gamma^\mu,\tilde I\}=0$ implies
\be\label{D12}
\cD^2=\partial^\mu\partial_\mu-m^2.
\ee
By using
\be\label{D13}
\big[\cK_m,\psi_\gamma(x)\big]=\tilde M_{\gamma\delta}\psi_\delta(x)
\ee
we find for the one-hole wave function $\hat q$ \eqref{H8a} the same evolution equation as for the one-particle wave function $q$, 
\be\label{D14} 
\cD\hat q=0.
\ee

It is instructive to write the one-particle evolution equation \eqref{D10a} in terms of the two-component complex Weyl spinor $\varphi$ defined by eq. \eqref{H14}
\be\label{D15}
\partial_t\varphi=\tau_k\partial_k\varphi-i\tau_2(m-i\tilde m)\varphi^*.
\ee
The fact that $\varphi^*$ appears shows that the  multiplication with $\tilde M$ is not compatible with the complex structure \eqref{186B}. In terms of $\chi$ \eqref{H19} and $\psi_M$ \eqref{H21} one finds
\ba\label{D16}
\partial_t\varphi&=&\tau_k\partial_k\varphi+(m-i\tilde m)\chi,\nn\\
\partial_t\chi&=&-\tau_k\partial_k\chi-(m+i\tilde m)\varphi,
\ea
and 
\ba\label{D17}
&&\partial_t\psi_M=\gamma^0_{(W)}(\gamma^k_{(W)}
\partial_k-im\bar\gamma_{(W)}+\tilde m)\psi_M,\nn\\
&&(i\gamma^\mu_{(W)}\partial_\mu+m\bar\gamma_{(W)}+i\tilde m)\psi_M=0,
\ea
with $\gamma^\mu_{(W)}$ the Dirac matrices in the Weyl representation \eqref{H22}. 

\bigskip\noindent
{\bf   2. General solution for single massive Majorana 

~particle}

\medskip
In order to find the general solution of eq. \eqref{D10a} we expand similar to eq. \eqref{H26}. We may consider modes with a fixed momentum $p$. (As familiar in quantum mechanics, these modes are not normalizable for infinite volume and may be considered as limiting cases of normalizable wave packets.) For every given $p\neq 0$ we can define 
\be\label{H29}
H(p)=\frac{p_k\tau_k}{|p|}~,~H^2(p)=1~,~p_kp_k=|p|^2,
\ee
such that $H(p)$ has eigenvalues $\pm 1$. Decomposing $\tilde\varphi(p)$ according to the eigenvalues of $H(p)$,
\be\label{30}
H(p)\tilde\varphi_\pm(p)=\mp\tilde\varphi_\pm(p),
\ee
and using
\ba\label{D19}
\tilde \chi_\pm(t,p)&=&-i\tau_2\tilde\varphi^*_\pm(t,-p),\nn\\
H(p)\tilde\chi_\pm(t,p)&=&\mp\tilde\chi_\pm(t,p),
\ea
one has   
\ba\label{D18}
\varphi(t,x)&=&\int_p\big(\tilde\varphi_+(t,p)e^{ipx}+
\tilde \varphi_-(t,p)e^{ipx}\big),\nn\\
\chi(t,x)&=&\int_p\big(\tilde \chi_+(t,p)e^{ipx}+\tilde\chi_-(t,p)e^{ipx}\big).
\ea
This yields, for $\tilde m=0$,
\ba\label{D20}
\partial_t\tilde\varphi_\pm(p)&=&\mp i|p|\tilde\varphi_\pm(p)+m\tilde\chi_\pm(p),\nn\\
\partial_t\tilde\chi_\pm(p)&=&\pm i|p|\tilde\chi_\pm(p)-m\tilde\varphi_\pm(p),
\ea
with general solution
\ba\label{D21}
\tilde\varphi_+(t,p)&=&b_+(p)e^{-i\omega t}-
\frac{\omega-|p|}{m}\tau_2b^*_+(-p)e^{i\omega t},\nn\\
\tilde\chi_+(t,p)&=&-i\tau_2b^*_+(-p)e^{i\omega t}-i
\frac{\omega-|p|}{m}b_+(p)e^{-i\omega t},\nn\\
\tilde\varphi_-(t,p)&=&b_-(p)e^{i\omega t}+
\frac{\omega-|p|}{m}\tau_2 b^*_-(-p)e^{-i\omega t},\\
\tilde\chi_-(t,p)&=&-i\tau_2 b^*_-(-p)e^{-i\omega t}+i
\frac{\omega-|p|}{m}b_-(p)e^{i\omega t},\nn
\ea
where
\be\label{D22}
\omega=\omega(p)=+\sqrt{|p|^2+m^2}.
\ee

\bigskip\noindent
{\bf 3. Classical probabilities for Majorana fermions}

\medskip
The general solution for $\varphi(t,x)$ depends on four complex functions contained in $b_\pm(p)$. From the real and imaginary part of $\varphi(t,x)$ we can infer the four real functions $q_\gamma(x)$ and therefore the classical probability distributions which describe the one-particle states. As an example, we consider the special solution $b_-(p)=0,b_{+,2}(p)=0,b_{+,1}(p)=f(p)$  real, for which
\ba\label{D23}
\varphi_1(t,x)&=&\int_pf(p)e^{i(px-\omega t)},\nn\\
\varphi_2(t,x)&=&\int_p-i\frac{\omega-|p|}{m}f(-p)e^{i(px+\omega t)},
\ea
or
\ba\label{D24}
q_1(t,x)&=&\int_p f(p)\cos (px-\omega t),\nn\\
q_2(t,x)&=&\int_p f(p)\sin (px-\omega t),\nn\\
q_3(t,x)&=&-\int_p f(p)\frac{\omega-|p|}{m}\sin(px-\omega t),\nn\\
q_4(t,x)&=&-\int_p f(p)\frac{\omega-|p|}{m}\cos(px-\omega t).
\ea
We may further take $f(p)=f\delta(p-\hat p)$ with $\hat p=(0,0,-P),P>0$, such that, with $\omega=\sqrt{P^2+m^2}$,
\ba\label{D25}
q_1(t,x)&=&f\cos(Px_3+\omega t)~,~q_2(t,x)=-f\sin (Px_3+\omega t),\nn\\
q_3(t,x)&=&f\frac{\omega-P}{m}\sin (Px_3+\omega t),\nn\\
q_4(t,x)&=&-f\frac{\omega-P}{m}
\cos (Px_3+\omega t).
\ea
For this particular example the normalization $\int_x q^2_\gamma(x)=1$ requires (with $V$ the space-volume)
\be\label{D26}
f^2=\frac{1}{2V}\frac{m^2}{m^2+P^2-P\sqrt{P^2+m^2}}.
\ee

For the general solution \eqref{D24} the probabilities $p_\gamma(t,x)=q^2_\gamma(t,x)$ describe the time dependent probability distribution of a classical statistical ensemble which accounts for all properties of a propagating massive single Majorana fermion. If the vacuum is the totally empty state $g_0=|0\rangle$ the states $(\gamma,x)$ correspond to ``Ising states'' $\tau$ where only one bit at $x$ of species $\gamma$ is occupied and all others are empty. (For a different $g_0$ the classical states associated to $(\gamma,x)$ are more complicated.) The probability of a state $(\gamma,x)$ is given by $p_\gamma(x)= q^2_\gamma(x)$, e.g. according to eq. \eqref{D24} with a proper normalization of $f(p)$. 

The probability distribution $\big \{p_\gamma(x)\big\}$ properly accounts for all characteristic quantum interference phenomena for modes with different momenta $p$. This extends to the correct description of interference for multi-particle states. (For a concrete example of a two-particle state for massless Majorana spinors cf. ref. \cite{CWMSa}.) The appearance of interference phenomena is not surprising in view of the formulation of the time evolution in terms of a Grassmann functional integral. On a deeper level, it is related to the fact that simple time evolution equations \eqref{35B} are linear in the classical wave function $q_\tau$, but not linear in the classical probabilities $p_\tau$.

\section{Dirac fermions}
\label{Diracfermions}

Dirac spinors can be composed of two different Majorana spinors with equal mass. The action reads
\ba\label{E1}
S&=&\int_{t,x}\big\{\psi_1(\partial_t-T_k\partial_k+m\gamma^0\tilde I)\psi_1\nn\\
&&+\psi_2(\partial_t-T_k\partial_k+m\gamma^0\tilde I)\psi_2\big\},
\ea
with $T_k,\tilde I,\gamma^0$ given by eqs. \eqref{18AA}, \eqref{18ABa}, \eqref{F7}. (For simplicity we take $\tilde m=0$ in eq. \eqref{D1a}.) The functional integral extends now over the Grassmann variables $\psi_{1,\gamma}(t,x)$ and $\psi_{2,\gamma}(x,t)$. For a free theory the action involves separate pieces for $\psi_1$ and $\psi_2$, $S=S_1+S_2$. If the initial state $g(t_0)=g_{in}$ factorizes into two factors, one involving only $\psi_1$ and the other only $\psi_2$, the Grassmann wave function $g(t)$ factorizes for all $t$. There is no need, however, for $g(t_0)$ to factorize and general states $g(t)$ will not be of a factorisable form.

\bigskip\noindent
{\bf  1. Complex structure}

\medskip
We introduce a complex structure by defining the four-component complex Grassmann variables $\psi_D$
\be\label{E2}
\psi_D=\psi_1+i\psi_2~,~\psi^*_D=\psi_1-i\psi_2.
\ee
In terms of the general discussion in sect. \ref{Complex structure} the matrices $K$ and $I$ act as
\be\label{E3}
K{\psi_1\choose \psi_2}={~~\psi_1\choose -\psi_2}~,~
I{\psi_1\choose\psi_2}={-\psi_2\choose ~~\psi_1},
\ee
such that $\psi_1$ is the ``real part'' of $\psi_D$, and $\psi_2$ the ``imaginary part''. The complex structure employed for Dirac spinors differs from the complex structure \eqref{186B}. We can express the action \eqref{E1} in terms of $\psi_D$ as
\ba\label{E4}
S&=&\int_{t,x}\psi^\dagger_D(\partial_t-T_k\partial_k+m\gamma^0
\tilde I)\psi_D\nn\\
&=&-\int_{t,x}\bar\psi_D(\gamma^\mu\partial_\mu-m\tilde I)\psi_D,
\ea
where we define 
\be\label{E5}
\bar\psi_D=\psi^\dagger_D\gamma^0.
\ee

The matrices $\gamma^\mu$ and $\tilde I$ appearing in eqs. \eqref{E4}, \eqref{E5} are all real. We may also introduce a purely imaginary matrix
\be\label{E6}
\bar\gamma=-i\gamma^0\gamma^1\gamma^2\gamma^3=-i\tilde I~,~\bar\gamma^2=1.
\ee
It is hermitean and anticommutes with all Dirac matrices $\gamma^\mu$,
\be\label{E7}
\bar\gamma^\dagger=\bar\gamma~,~\{\bar\gamma,\gamma^\mu\}=0~,~
[\bar\gamma,\Sigma^{\mn}]=0.
\ee
Often $\bar\gamma$ is denoted as $\gamma^5$. On the level of the eight variables $\psi_{1,\gamma}$ and $\psi_{2,\gamma}$ the matrix $\bar\gamma$ is defined as a real $8\times 8$ matrix, with $I_2$ not acting on the index $\gamma$, 
\be\label{E8}
\bar\gamma=-I_2\otimes\tilde I~,~I_2={0,-1\choose 1,~~0}. 
\ee
(The matrix $I$ defining the complex structure is given by $I_2$ according to eq. \eqref{E3}.)

Since $\bar\gamma$ commutes with the Lorentz-generators $\Sigma^{\mn}$ the projections
\be\label{E8A}
\psi_L=\frac12(1+\bar\gamma)\psi_D~,~\psi_R=\frac12(1-\bar\gamma)\psi_D
\ee
are representations of the Lorentz group. They transform as the two inequivalent fundamental spinor representations and describe Weyl spinors. In terms of $\psi_{1,\gamma},\psi_{2,\gamma}$ one has
\be\label{E8B}
\psi_{L,R}=\frac12
\left(\begin{array}{c}
\psi_{1,1}\mp\psi_{2,2}+i\psi_{2,1}\pm i\psi_{1,2}\\
\psi_{1,2}\pm\psi_{2,1}+i\psi_{2,2}\mp i\psi_{1,1}\\
\psi_{1,3}\mp\psi_{2,4}+i\psi_{2,3}\pm i\psi_{1,4}\\
\psi_{1,4}\pm\psi_{2,3}+i\psi_{2,4}\mp i\psi_{1,3}
\end{array}\right),
\ee
where the upper sign relates to $\psi_L$ and the lower to $\psi_R$. We observe that both $\psi_L$ and $\psi_R$ receive contributions both from $\psi_1$ and $\psi_2$. They differ from the Weyl spinors $\zeta_1,\zeta_2$ that can be formed from $\psi_1$ and $\psi_2$ according to eq. \eqref{71LA}. 

Using $\bar\gamma$ we can write the action \eqref{E4} as
\ba\label{E9}
S&=&-\int_{t,x}\bar\psi_D(\gamma^\mu\partial_\mu- im\bar\gamma)\psi_D=-iS_M,\nn\\
S_M&=&-\int_{t,x}\bar\psi_D(i\gamma^\mu\partial_\mu+m\bar\gamma)\psi_D.
\ea
Defining the hermitean conjugation as a map $\psi_D\to \psi^*_D$, accompanied by a complex conjugation of all coefficients in the Grassmann algebra and a total transposition (reordering) of all Grassmann variables, the hermiticity of the Minkowski action, $S^\dagger_M=S_M$, is easily verified. 

\bigskip\noindent
{\bf  2. Charge and electromagnetic fields}

\medskip 
The action \eqref{E1} is invariant under all symmetry transformations defined in the preceding sections, i.e. those leaving $S_1$ or $S_2$ separately invariant. There are further symmetries related to transformations between $\psi_1$ and $\psi_2$. Rotations between $\psi_1$ and $\psi_2$ can be accounted for by the infinitesimal transformation
\be\label{E10}
\delta\tilde\psi=\delta{\psi_1\choose\psi_2}=-\beta I_2\tilde \psi=
{~~\beta\psi_2\choose -\beta\psi_1}.
\ee
The discretization is the same for $\psi_1$ and $\psi_2$. It therefore respects the invariance with respect to the transformation \eqref{E10}. This rotation can be transferred to a phase rotation of the complex Dirac spinor
\be\label{E11}
\delta\psi_D=-i\beta\psi_D.
\ee
We associate the transformations \eqref{E10}, \eqref{E11} with the $U(1)$-transformations of electromagnetism. The Dirac spinor carries charge $Q=-1$ and can be identified with the electron. Both components $\psi_L$ and $\psi_R$ \eqref{E8A} transform as Weyl spinors with the same electric charge $Q$. 

We may extend the action by considering Dirac spinors in an external electromagnetic field $A_\mu(t,x)$. As usual this is done by replacing the derivatives $\partial_\mu$ in eq. \eqref{E9} by covariant derivatives $D_\mu=\partial_\mu+ieA_\mu$. This adds to the action a piece
\be\label{E12}
\Delta S=-ie\int_{t,x}\bar\psi_D\gamma^\mu A_\mu\psi_D.
\ee
In terms of the real Grassmann algebra involving $\psi_1$ and $\psi_2$ the additional piece reads
\be\label{E13}
\Delta S=-e\int_{t,x}\big\{\psi_1(A_0-A_kT_k)\psi_2
-\psi_2(A_0-A_kT_k)\psi_1\big\}.
\ee
For discrete time steps the first Grassmann variable is taken at $t$ and the second at $t+\epsilon$. In the continuum limit the two terms in eq. \eqref{E13} are equal. We may consider the fields $A_\mu$ as ``sources'' for fermion bilinears involving $\psi_1$ and $\psi_2$. For nonzero $A_\mu$ the action is no longer a sum of independent pieces $S_1+S_2$ for $\psi_1$ and $\psi_2$, respectively.

\bigskip\noindent
{\bf  3. Weyl representation}

\medskip
One may use the similarity transformation \eqref{67A}
\be\label{E14}
\psi_{D,W}=A^{-1}\psi_D~,~
\gamma^\mu_{(W)}=A^{-1}\gamma^\mu_{(M)}A,
\ee 
in order to bring the Dirac matrices to the Weyl representation \eqref{H22}. The action \eqref{E9}, \eqref{E12} retains its form, with the replacements $\gamma^\mu\to\gamma^\mu_{(W)},~\psi_D\to\psi_{D,W},\bar\psi_D\to\psi^\dagger_{D,W}\gamma^0_{(W)}=\bar\psi_{D,W}$. In the Weyl representation the matrix $\bar \gamma$ is diagonal, $\bar\gamma =diag(1,1,-1,-1)$. The Dirac spinor $\psi_{D,W}$ can therefore be written in terms of two-component spinors $\psi_L$ and $\psi_R$,
\ba\label{E15}
\psi_{D,W}&=&{\psi_L\choose \psi_R}~,~\psi_L=(\zeta_1+i\zeta_2)~,~\psi_R=
(\xi_1+i\xi_2),\nn\\
\zeta_a&=&\frac{1}{\sqrt{2}}
{\psi_{a,1}+i\psi_{a,2}\choose \psi_{a,3}+i\psi_{a,4}},\nn\\
\xi_a&=&\frac{1}{\sqrt{2}}
{-\psi_{a,3}+i\psi_{a,4}\choose ~~\psi_{a,1}-i\psi_{a,2}}=-i\tau_2\zeta^*_a,
\ea
which, in turn, can be written as linear combinations of $\zeta_1$ and $\zeta_2$ or $\xi_1$ and $\xi_2$. 

We observe that in this basis the operation of complex conjugation is represented by an involution $\hat K$ different from $K$ in eq. \eqref{E3}. In terms of the eight-component field $\tilde \psi=(\psi_1,\psi_2)$ one has $\hat K=diag (1,-1,1,-1,-1,1,-1,1)$. The matrix $I$ remains $I_2$ as in eq. \eqref{E3}, and $I_2$ and $\hat K$ anticommute
\be\label{E16}
\hat K={\tilde K,~~0\choose \hspace{0.15cm}0,-\tilde K}~,~
I_2={0,-1\choose 1,~~0}~,~\{\hat K,I_2\}=0.
\ee
Thus the similarity transformation \eqref{E14} changes the complex structure. On the level of a real representation it can be expressed by real $8\times 8$ matrices $A'$ such that a similarity transformation may, in principle, transform both $K$ and $I$. The matrix $I$ remains unchanged, since $iA\psi=A(i\psi)$ in the complex representation. For $A\neq A^*$, however, the matrix $K$ changes since $(A\psi)^*\neq A\psi^*$. 

In the Weyl basis \eqref{E15} we can take $\psi_L$ and $\psi_R$ as two-component complex spinors. In terms of $\psi_L$ and $\psi_R$ the action reads
\ba\label{E17}
S&=&-\int_{t,x}\big\{\bar\psi_L\gamma^\mu(\partial_\mu+ieA_\mu)\psi_L+\bar\psi_R
\gamma^\mu(\partial_\mu+ieA_\mu)\psi_R\nn\\
&&-im(\bar\psi_R\psi_L-\bar\psi_L\psi_R)\big\},
\ea
where we define
\ba\label{E18}
\bar\psi&=&(\bar\psi_R,\bar\psi_L)=-i(\psi^\dagger_R~,~\psi^\dagger_L)\nn\\
\bar\psi_L&=&\bar\psi\left(\frac{1-\bar\gamma}{2}\right)~,~\bar\psi_R=\bar\psi
\left(\frac{1+\bar\gamma}{2}\right).
\ea
For $m=0$ it is invariant under separate ``chiral phase rotations'' of $\psi_L$ and $\psi_R$. The transformation $\delta\psi_L=i\alpha\psi_L,\delta\psi_R=-i\alpha\psi_R$ corresponds to $\delta\zeta_a=i\alpha\delta\zeta_a$ and therefore to a simultaneous transformation of the type \eqref{71B} for $\psi_1$ and $\psi_2$. 

\bigskip\noindent
{\bf 4. Left-handed representation}

\medskip
We can also reformulate the action in terms of two left handed Weyl spinors $\psi_L$ and $\psi^c_L$ by introducing
\ba\label{E19}
\psi^c_L&=&i\tau_2\psi^*_R=\zeta_1-i\zeta_2,\nn\\
\psi_R&=&-i\tau_2(\psi^c_L)^*~,~\bar\psi_R=(\psi^c_L)^T\tau_2.
\ea
This yields
\ba\label{E20}
S&=&\int_{t,x}\Big\{\psi^\dagger_L\big[\partial_t+ieA_0
-\tau_k(\partial_k+ieA_k)\big]\psi_L\nn\\
&&+(\psi^c_L)^\dagger\big[\partial_t-ieA_0-\tau_k(\partial_k-ieA_k)\big]\psi^c_L\nn\\
&&+im\big[(\psi^c_L)^T\tau_2\psi_L+(\psi^c_L)^\dagger\tau_2\psi^*_L\big]\Big\}.
\ea
The charge of $\psi^c_L$ is opposite to $\psi_L$ - if $\psi_L$ describes a left-handed electron, $\psi^c_L$ accounts for the left-handed positron.

We may group $\psi_L$ and $\psi^c_L$ into a four-component complex spinor
\be\label{269A}
\psi_{LL}={\psi_L\choose \psi^c_L}={\zeta_1+i\zeta_2\choose \zeta_1-i\zeta_2}.
\ee
In this representation the matrix $\tilde I$ defined by eq. \eqref{18ABa} is represented by a multiplication with $i$,
\be\label{269B}
\psi_{LL}[\tilde I\tilde \psi]=i\psi_{LL},
\ee
while the matrix $I_2$, which acts on $(\zeta_1,\zeta_2)$ or $(\psi_1,\psi_2)$ as defined in eq. \eqref{E8}, is now represented as 
\be\label{269C}
\psi_{LL}[I_2\tilde\psi]={i,~~0\choose  0,-i}\psi_{LL}.
\ee
We observe that the role of $\tilde I$ and $I_2$ is exchanged if we compare the complex spinors $\psi_{LL}$ and $\psi_{D,W}$,
\ba\label{269D}
\psi_{D,W}[I_2\tilde\psi]&=&i\psi_{D,W},\\
\psi_{D,W}[\tilde I\tilde\psi]&=&{i,~~0\choose 0,-i}\psi_{D,W}
=i\bar\gamma\psi_{D,W}.\nn
\ea
The complex structures associated to $\psi_{LL}$ and $\psi_{D,W}$ are therefore different. Nevertheless, the operation of complex conjugation of $\psi_{LL}$ or $\psi_{D,W}$ corresponds in the real representation $\tilde\psi$ to the same matrix multiplication with $\hat K=diag(\tilde K,-\tilde K)$. 

\bigskip\noindent
{\bf  5. Enhanced symmetries for free massless Dirac spinors}

\medskip
For $m=0,A_\mu=0$ the free theory exhibits a symmetry of $SU(2)$-rotations among $\psi_L$ and $\psi^c_L$,
\be\label{E21}
\delta\psi_{LL}=i\omega_j\hat \tau_j\psi_{LL},
\ee
with $\omega_3=\beta$ identified with the transformation \eqref{E11} and $\hat\tau_j$ the Pauli matrices acting on the two components of $\psi_{2d}$ (not on the spinor components of $\psi_L$ and $\psi^c_L$). For example, the transformation with $\omega_1$ acts as
\ba\label{E22}
\delta\psi_L&=&-\omega_1\tau_2\psi^*_R~,~\delta\psi_R=-\omega_1\tau_2\psi^*_L,\nn\\
\delta\zeta_1&=&i\omega_1\zeta_1~,~\delta\zeta_2=-i\omega_1\zeta_2. 
\ea
This corresponds to a transformation of the type \eqref{71B}, now with opposite directions for $\psi_1$ and $\psi_2$. The situation is similar for the transformation $\sim\omega_2$, where
\ba\label{E23}
\delta\zeta_1&=&-i\omega_2\zeta_2~,~\delta\zeta_2=-i\omega_2\zeta_1,\nn\\
\delta\psi_1&=&-\omega_2\tilde I\psi_2~,~\delta\psi_2=-\omega_2\tilde I\psi_1
\ea
leads again to a transformation among $\psi_{1,\gamma}$ and $\psi_{2,\gamma}$ which involves $\tilde I$.

\newpage\noindent
{\bf  6. Evolution equation for Dirac fermions}  

\medskip
For the action $S+\Delta S$ \eqref{E1}, \eqref{E13} the Grassmann evolution equation $\partial_t g={\cal K}g$ involves ${\cal K}={\cal K}_0+\cK_m+\Delta{\cal K}$
\ba\label{X1}
{\cal K}_0+\cK_m&=&\int_x\sum_{a=1,2}\frac{\partial}{\partial\psi_a(x)}
(T_k\partial_k-m\gamma^0\tilde I)\psi_a(x),\nn\\
\Delta{\cal K}&=&e\int_x\left[\frac{\partial}{\partial\psi_1(x)}
(A_0(x)-A_k(x) T_k)\psi_2(x)\right.\nn\\
&&\left.-\frac{\partial}{\partial\psi_2(x)}
(A_0(x)-A_k(x)T_k)\psi_1(x)\right].
\ea
Here we work in the Majorana basis \eqref{18AA}, \eqref{18ABa}, \eqref{F7} and we have suppressed the spinor index $\gamma$. The evolution equation 
\be\label{231}
\partial_t g(t)=\cK g(t)=(\cK_0+\cK_m+\Delta\cK)g(t)
\ee
describes the dynamics for an arbitrary number of charged relativistic fermions (and their antiparticles) in external electromagnetic fields. 

The evolution equation for the classical wave function \eqref{P35}, and therefore for the probability distribution, obtains from eq. \eqref{P44}. In summary, we have formulated a time evolution equation for a classical wave function $q_\tau(t)$ which describes an arbitrary number of charged electrons in external electric and magnetic fields. In the next section we will see that its restriction to one-particle states yields the relativistic Dirac equation. As usual, a non-relativistic approximation will yield the familiar Schr\"odinger equation for a particle in a potential or moving in external magnetic fields.

\medskip\noindent
{\bf 7. Classical observables}

Classical observables can be constructed from linear combinations of products of occupation numbers. We combine the index $a=1,2$ for the two Majorana spinors $\psi_a$ with the index $\gamma$ into a common index $\epsilon=(\gamma,a),\epsilon=1\dots 8$. The Grassmann operators corresponding to the occupation numbers $N_\epsilon(x)$ read then 
\be\label{233A}
\cN_\epsilon(x)=\frac{\partial}{\partial\psi_\epsilon(x)}\psi_\epsilon(x).
\ee
In a formulation with discrete lattice points they obey 
\be\label{FF1}
\cN^2_\epsilon(x)=\cN_\epsilon(x)~,~\big[\cN_\epsilon(x),\cN_\eta(y)\big]=0.
\ee
The basis elements $g_\tau$ are eigenstates of $\cN_\epsilon(x)$, 
\be\label{FF3}
\cN_\epsilon(x)g_\tau=\big(N_\epsilon(x)\big)_\tau g_\tau,
\ee
with $\big(N_\epsilon(x)\big)_\tau=0$ if $g_\tau$ contains a factor $\psi_\epsilon(x)$, and $\big(N_\epsilon(x)\big)_\tau=1$ if no such factor is present. The expectation value of a product of occupation numbers obeys
\ba\label{FF4}
&&\kl N_{\epsilon_1}(x_1)N_{\epsilon_2}(x_2)\dots N_{\epsilon_m}(x_m)\kr\nn\\
&&=\int\cD\psi(t)\tilde g(t)\cN_{\epsilon_1}(x_1)\dots \cN_{\epsilon_m}(x_m)g(t)\nn\\
&&=\sum_\tau p_\tau(t)\big(N_{\epsilon_1}(x_1)\big)_\tau\dots\big(N_{\epsilon_m}(x_m)\big)_\tau,
\ea
corresponding to the standard rule for classical statistics. Obviously, the expectation values of classical observables can be computed from the probability distribution $p_\tau(t)$ and do not involve the signs $s_\tau$ in eq. \eqref{35A}. Classical products of classical observables commute.

In order to compute the time evolution of expectation values of classical observables we define
\be\label{FF5}
{\cal M}_{\epsilon\eta}(x)=\frac{\partial}{\partial\psi_\epsilon(x)}\psi_\eta(x),
\ee
and use the relation
\ba\label{FF6}
\big[{\cal M}_{\epsilon\eta}(x),\cK\big]&=&\sum_\alpha\big\{(T_k)_{\epsilon\alpha}\partial_k\
\frac{\partial}{\partial\psi_\alpha(x)}\psi_\eta(x)\nn\\
&&+(T_k)_{\eta\alpha}\frac{\partial}{\partial\psi_\epsilon(x)}\partial_k\psi_\alpha(x)\\
&&+W_{\epsilon\alpha}(x){\cal M}_{\alpha\eta}(x)-{\cal M}_{\epsilon\alpha}(x)W_{\alpha\eta}(x)\big\},\nn
\ea
with 
\be\label{FF7}
W_{\epsilon\eta}(x)=-m(\gamma^0\tilde I)_{\alpha\beta}
\delta_{ab}+e\big(A_0(x)\delta_{\alpha\beta}-A_k(x)(T_k)_{\alpha\beta}\big)\epsilon_{ab},
\ee
and $\partial_k=\partial/\partial x_k,\epsilon_{ab}=-\epsilon_{ba},\epsilon_{12}=1,\eta=(\beta,b),(T_k)_{\epsilon\eta}=(T_k)_{\alpha\beta}\delta_{ab}=(T_k)_{\eta\epsilon}$. We observe that $W$ is antisymmetric, $W_{\epsilon\eta}(x)=-W_{\eta\epsilon}(x)$. A local particle number can be defined as 
\be\label{FF8}
\cN(x)=\sum_\epsilon\cN(x)=\sum_\epsilon{\cal M}_{\epsilon\epsilon}(x),
\ee
and obeys
\be\label{FF9}
\big[\cN(x),\cK \big]=(T_k)_{\eta\alpha}\partial_k{\cal M}_{\alpha\eta}.
\ee
According to eq. \eqref{95D} the time evolution of the classical mean local particle number involves expectation values of ``off-diagonal'' operators ${\cal M}_{\alpha\eta}$
\be\label{FF10}
\partial_t\kl N(x)\kr=(T_k)_{\eta\alpha}\partial_k\kl {\cal M}_{\alpha\eta}\kr.
\ee
The particle number 
\be\label{FF11}
\cN=\int_x\cN(x)~,~[\cN,\cK]=0,
\ee
is conserved.

\section{Quantum mechanics for particle in a potential}
\label{Quantum mechanics for particle in a potential}

\noindent
{\bf 1. Dirac equation}

We next concentrate on one-particle states, described by Grassmann elements
\be\label{X2}
g_1(t)=\int_x\left(q_{1,\gamma}(t,x)
\frac{\partial}{\partial\psi_{1,\gamma}(x)}+q_{2,\gamma}(t,x)
\frac{\partial}{\partial\psi_{2,\gamma}(x)}\right)g_0.
\ee
Here $g_0$ is some arbitrary static ``vacuum state''. For ${\cal K}g_0=0$ we find for the one-particle wave function the Dirac equation
\ba\label{X3}
&&(\gamma^\mu\partial_\mu -m\tilde I) q_1=eA_\mu\gamma^\mu q_2,\nn\\
&&(\gamma^\mu\partial_\mu -m\tilde I) q_2=-e A_\mu\gamma^\mu q_1.
\ea

For the derivation of the Dirac equation we use
\ba\label{X4}
&&\left[\Delta{\cal K}~,~\frac{\partial}{\partial\psi_{1,\gamma}(x)}\right]\nn\\
&&=-e\left(A_0(x)\frac{\partial}{\partial\psi_{2,\gamma}(x)}-A_k(x)(T_k)_{\delta\gamma}\frac{\partial}{\partial\psi_{2,\delta}(x)}\right)\nn\\
&&\left[ \Delta{\cal K},\frac{\partial}{\partial\psi_{2,\gamma}(x)}\right] \\
&&=e\left(A_0(x)\frac{\partial}{\partial\psi_{1,\gamma}(x)}-A_k(x)(T_k)_{\delta\gamma}\frac{\partial}{\partial\psi_{1,\delta}(x)}\right).\nn
\ea
The time evolution equation for the one particle wave function then extends eq. \eqref{H11} to 
\ba\label{276A}
\partial_t q_1&=&(T_k\partial_k -m\gamma^0\tilde I) q_1+e(A_0-A_k T_k)q_2,\nn\\
\partial_tq_2&=&(T_k\partial_k -m\gamma^0\tilde I) q_2-e(A_0-A_kT_k)q_1.
\ea
This is equivalent to the Dirac equation \eqref{X3} which can also be written as a matrix equation. 
\ba\label{276B}
&&(\gamma^\mu D_\mu-m\tilde I){q_1\choose q_2}=0,\nn\\
&&D_\mu= \partial_\mu+eA_\mu
{0,-1\choose 1,~~0}=\partial_\mu+eA_\mu I_2.
\ea
The usual complex form of the Dirac equation is recovered if we use a complex one-particle wave function
\be\label{276C}
\varphi_D=q_1+iq_2~,~\gamma^\mu(\partial_\mu+ieA_\mu)\varphi_D=im\bar\gamma\varphi_D.
\ee
This equation holds for an arbitrary representation of the Dirac matrices $\gamma^\mu$. 

The derivation of the Dirac equation for a one-particle state \eqref{X2} has only used the general evolution equation for Grassmann elements $g(t)$ and the condition $\cK g_0=0$. It therefore describes the dynamics for a very extended family of classical probability distributions or classical wave functions. Indeed, it is sufficient that $g_0$ is an arbitrary static state (not necessarily a priori with a fixed particle number). This reflects the physical property that isolated one-particle states can occur under a wide variety of circumstances, and that for sufficient isolation the properties of the environment do not matter for the dynamics of the isolated particle. In this context we emphasize, however, that our model only describes external electromagnetic fields while we do not account for the fields generated by the particles that may be present  in the environment. In this sense our setting describes ``real physics'' only in a situation where the electromagnetic fields generated by all present particles can be approximated by a ``mean field'' that is independent of individual particle positions. This is precisely the setting of standard one-particle quantum mechanics. 

The presence of electromagnetic fields influences the conditions for a static ``vacuum'' or ``environment'', $\cK g_0=0$. We will concentrate here on simple vacuum states as $g_0=1$ or $g_0=|0\kr$, where $|0\kr$ involves a product of all Grassmann variables and obeys $\cN|0\kr=0$. These states are static for arbitrary electromagnetic fields. For simplicity we choose $g_0=|0\kr$ such that 
\be\label{VV1}
\cN g_1=g_1.
\ee
(The following discussion can be easily generalized, however, to all static states $g_0$ which are eigenvalues of $\cN,\cN g_0=m\gamma_0$, by using a modified local occupation number where a suitable constant is subtracted from $\cN(x)$.)

\medskip\noindent
{\bf 2. Particle observables}

Let us consider the expectation value of the local occupation number 
\be\label{VV1A}
\cN(x)=\sum_\epsilon \cN_\epsilon(x).
\ee
We use the identity
\be\label{VV2}
\left[ \cN(x),\int_y q_\epsilon(t,y)\frac{\partial}{\partial\psi_\epsilon(y)}\right]=
\sum_\epsilon q_\epsilon(t,x)\frac{\partial}{\partial\psi_\epsilon(x)}
\ee
in order to establish the simple relation 
\be\label{VV3}
\kl \cN(x)\kr=\sum_\epsilon q^2_\epsilon(t,x).
\ee
This holds for a wave function $g(t)=g_1(t)$, and we use eq. \eqref{G19} employing appropriate basis elements (a subset of $\{g_\tau\}$)
\ba\label{VV4}
&&g_{1,\epsilon}(x)=\frac{\partial}{\partial\psi_\epsilon(x)}|0\kr~,~\tilde g_{1,\eta}(y)=\psi_\eta (y),\nn\\
&&\int \cD\psi\tilde g_{1,\eta}(y)g_{1,\epsilon}(x)=\delta_{\eta\epsilon}\delta(x-y),
\ea
with 
\be\label{VV5}
g_1=\int_x q_\epsilon(x,t)g_{1,\epsilon}(x)~,~\tilde g_1=\int_y q_\eta(y,t)\tilde g_{1,\eta}(y).
\ee

The normalization $\int\cD\psi\tilde g_1g_1=1$ implies for the one-particle wave function the normalization
\be\label{VV6}
\int_x \sum_\epsilon\big(q_\epsilon(x)\big)^2=\int_x\varphi^\dagger_D(x)\varphi_D(x)=1,
\ee
which is compatible with $\kl\cN\kr=1$, cf. eq. \eqref{VV1}. The normalization \eqref{VV6} is preserved by the unitary time evolution of the one-particle wave function. Indeed, we can write the Dirac equation \eqref{276C} as a Schr\"odinger-type equation 
\be\label{VV7}
i\hbar\partial_t\varphi_D=H\varphi_D,
\ee
with hermitean Hamiltonian $(T_k=\gamma^0\gamma^k)$
\be\label{VV8}
H=i\hbar T_k\partial_k+\hbar m\gamma^0\bar\gamma+\hbar e(A_0 -T_k A_k)=H^\dagger.
\ee

For a pure one-particle state the expectation value $\kl N(x)\kr$ can be interpreted in a natural way as the probability density to find the particle at the position $x$. This is precisely the standard interpretation of $\psi^\dagger_D(x)\psi_D(x)$ in one-particle quantum mechanics,
\ba\label{VV9}
&&w(x)=\kl N(x)\kr=\varphi^\dagger_D(x)\varphi_D(x),\nn\\
&&\int_xw(x)=1.
\ea
Since $N(x)$ is a classical observable, we can define the position of the particle as a classical observable 
\be\label{VV10}
X=\int_x xN(x).
\ee
The expectation value in classical statistics coincides with the standard quantum mechanics rule
\be\label{VV11}
\kl X\kr=\kl \int_x xN(x)\kr=\int_xxw(x)=\int_x\varphi^\dagger_D(x)x\varphi_D(x).
\ee

This finds a natural extension to arbitrary functions of the position observable
\be\label{VV12}
f(X)=\int_xf(x)N(x).
\ee
The expectation values of this type of classical observables follow again the rule of quantum mechanics
\be\label{VV13}
\kl f(X)\kr=\int_x\varphi^\dagger_D(x)f(x)\varphi_D(x).
\ee
In particular, one obtains the same formula for the dispersion $\kl X_kX_k\kr-\kl X_k\kr\kl X_k\kr$ as in quantum mechanics. We conclude that measurements of the position of a particle, or more generally the distribution of positions in an ensemble of one-particle states, can be described equivalently in a classical statistical ensemble with classical observables, or in quantum mechanics with Hamiltonian \eqref{VV8}. The complete time evolution of the distribution of positions is identical in both descriptions. This covers, in particular, the characteristic quantum interference in a double slit experiment. The time evolution \eqref{P35}, \eqref{P44}, \eqref{231} for the classical wave function and associated classical probability distribution $\{p_\tau\}$ produces for one-particle states exactly the quantum mechanical interference pattern.

\medskip\noindent
{\bf 3. Schr\"odinger equation}

Standard quantum mechanics for an electron in a potential is recovered from the non-relativistic approximation to the Dirac equation. This is well known, and we sketch here for completeness only the case $A_k=0$. The nonrelativistic approximation becomes valid if $eA_0$ and $iT_k\partial_k$ are small compared to $m$. Since $(\gamma^0\bar \gamma)^2=1$, it is convenient to choose a basis where $\gamma^0\bar\gamma=diag (1,1,-1,-1)$. With $V(x)=\hbar eA_0(x)$ and $M=\hbar m$ the Hamiltonian \eqref{VV8} takes the form
\be\label{VV14} 
H=
\left(\begin{array}{ccc}M&,&\sigma_kp_k\\\sigma^\dagger_kp_k&,&-M\end{array}\right)
+V(x),
\ee
where we use the standard quantum mechanical momentum operator
\be\label{VV15}
p_k=-i\hbar \partial_k.
\ee
The matrices
\be\label{VV16}
\sigma_1=-1~,~\sigma_2=i\tau_2~,~\sigma_3=i\tau_3
\ee
arise from 
\be\label{VV17}
T'_k=U^\dagger T_kU=-\left(\begin{array}{ccc}
0&,&\sigma_k\\\sigma^\dagger_k&,&0\end{array}\right),
\ee
with 
\be\label{VV18}
U=\frac{1}{\sqrt{2}}
\left(\begin{array}{ccc}\tau_1&,&-\tau_2\\-\tau_2&,&\tau_1\end{array}\right)~,~
U^\dagger U=1~,~U^\dagger=U.
\ee
(Note $T'_1=T_1,~T'_2=-T_2$.) The unitary matrix $U$ diagonalizes $\gamma^0\bar\gamma$, 
\be\label{VV19}
\gamma^0\bar\gamma=
\left(\begin{array}{ccc}0&,&-i\tau_3\\i\tau_3&,&0\end{array}\right)~,~
U^\dagger\gamma^0\bar\gamma U=
\left(\begin{array}{ccc}1&,&0\\0&,&-1\end{array}\right).
\ee
We observe for the matrices $\sigma_k$ the relations
\be\label{VV20}
\sigma^\dagger_k\sigma_l+\sigma^\dagger_l\sigma_k=\sigma_k\sigma^\dagger_l+\sigma_l\sigma^\dagger_k=2\delta_{kl},
\ee
which guarantee $\{T'_k,T'_l\}=2\delta_{kl}$.

We next decompose $\varphi_D$ into two-component wave functions, $\varphi^T_D=(\chi^T,\rho^T)$, which obey 
\ba\label{VV21}
i\hbar\partial_t\chi&=&(M+V)\chi+\sigma_kp_k\rho,\nn\\
i\hbar\partial_t\rho&=&(-M+V)\rho+\sigma^\dagger_kp_k\chi.
\ea
For the non-relativistic electron we consider the approximate solution $\rho=A\chi$, where $A=\sigma^\dagger_kp_k/(2M)$ is determined by requiring in leading order $\partial_t\rho=A\partial_t\chi$. Insertion into eq. \eqref{E25} yields for $\psi=\exp(iMt/\hbar)\chi$ the standard Schr\"odinger equation for a particle in a potential $V$,
\be\label{VV22}
i\hbar\partial_t\psi=\left(\frac{p_kp_k}{2M}+V\right)\psi.
\ee
As usual, the Dirac equation can also describe non-relativistic positrons. For this purpose one considers a second class of solutions $\chi=B\rho$, with $B=-\sigma_kp_k/(2M)$. For $\tilde \psi=\exp(iMt/\hbar)\rho^*$ one obtains
\be\label{VV23}
i\hbar\partial_t\tilde\psi=
\left(\frac{p_kp_k}{2M}-V\right)\tilde\psi,
\ee
and we note the change of sign of the potential due to the opposite charge of the positron.

All quantum mechanical phenomena extracted from solutions of the Schr\"odinger equation are described by our time evolution equation for a classical statistical ensemble of Ising-spins on a lattice. For a potential realizing a double-slit situation the standard interference pattern is predicted for this classical statistical ensemble to appear behind the slits. This holds provided that the initial state at some time $t_0$ corresponds to a one-particle state describing a particle moving towards the slits. 

\newpage\noindent
{\bf 4. Particle-wave duality}

The discreteness of measurement values in quantum mechanics can be traced back to the discrete occupation numbers of the Ising-type model. In quantum mechanics we may define an ``interval observable'' $J_{{\cal R}}$ by a function
\be\label{VV24}
J_\cR(x)=\left\{\begin{array}{ll}
1&\text{if}~x\in \cR\\0&\text{otherwise}
\end{array}\right).
\ee
It has the property
\be\label{VV25}
J^2_\cR=J_\cR,
\ee
such that its spectrum consists of the discrete values $0$ and $1$. According to the rules of quantum mechanics, the possible outcomes of a measurement of $J_\cR$ are $0$ or $1$. The interpretation in quantum mechanics is simple: either the particle is within the region (interval) $J_\cR$, in which case the measurement value $J_\cR=1$ will be found, or it is outside this region, and $J_\cR=0$ will be found. Particles are discrete objects - they are either inside or outside an interval. 

In our classical statistical Ising-type setting $J_\cR$ is a classical observable, given by 
\be\label{VV26}
J_\cR=\int_\cR N(x)=\sum_\cR N_L(x).
\ee
The sum $\sum_\cR$ extends over all lattice points within the region $\cR$. The classical observable $N_L(x)$ corresponds to a normalization of occupation numbers for the discrete lattice where $\big (N_{L,\epsilon}\big)_\tau=0,1$. The sum over species $N_L(x)=\sum\limits_\epsilon N_{L,\epsilon}(x)$, cf. eq. \eqref{VV1A}, can therefore take in any classical state $\tau$ only the discrete values $\big(N_L(x)\big)_\tau=(0,1\dots, 8)$, according to the total number of eight species. In consequence, for any state $\tau$ of the classical statistical ensemble, $(J_\cR)_\tau$ is a positive integer or zero. According to the standard rule of classical statistics these integers describe the possible outcomes of measurements. 

For a one particle state the total particle number equals one,
\be\label{VV27}
1=\int\limits_V N(x)=\sum_VN_L(x),
\ee
where $\sum\limits_V$ extends now over all lattice points in the total volume. Since $\cR$ must be contained in $V$ the maximal allowed value for $J_\cR$ in a one-particle state is one, such that $(J_\cR)_\tau=0,1$ are the only possible values of the classical observable for such a state. This is precisely the quantum mechanical rule. No new postulate is necessary for this measurement in quantum mechanics - the quantum rule is inferred from the standard rule of classical statistics. Of course, the expectation value of $J_\cR$ for a one particle state is the same in the quantum mechanical and the classical statistical description 
\be\label{VV28}
\kl J_\cR\kr=\int_x\varphi^\dagger_D(x)J_\cR(x)\varphi_D(x)=
\int_\cR\varphi^\dagger_D(x)\varphi_D(x).
\ee

While our model connects the discrete particle aspects in quantum  mechanics directly to the discrete classical Ising-spins, the continuous wave aspects also arise in a natural way. The quantum wave function is continuous because probability distributions and associated classical wave functions are continuous (at least piecewise). As we have mentioned already, the characteristic interference effects for waves and the superposition arise from the linearity of the fundamental evolution equation in the classical wave function.

\section{Conclusions}
\label{Conclusions}
We have derived the Schr\"odinger equation for a quantum particle in a potential from a classical statistical ensemble for Ising-spins. The dynamics of the classical statistical system has to be specified by an  appropriate evolution equation for the probability distribution. For this purpose we employ the classical wave function which is defined as the positive or negative root of the probability distribution. The proposed evolution equation is a linear differential equation for the classical wave function. The wave function at time $t$ obtains from the wave function at $t'$ by a rotation - this guarantees the preservation of the normalization of the probability distribution. 

We have exploited a map between the classical wave function and a Grassmann wave function which is an element of a real Grassmann algebra. In turn, the time evolution of the Grassmann wave function can be associated to a Grassmann functional integral. This allows us to formulate the evolution equation for the classical wave function in terms of the action of a functional integral. The symmetries of the  model, as Lorentz symmetry and electromagnetic gauge symmetry for our model of Dirac spinors in an external electromagnetic field, can be easily implemented in this way. Since the classical wave function is real we have to formulate the model in terms of a real Grassmann algebra.

Our model is regularized on a lattice of space points. On the one hand, this guarantees that mathematical expressions are well defined for a finite number of lattice points, with continuum limit of an infinite number of points taken at the end. On the other hand, the concept of classical Ising-spins or associated occupation numbers at every  point $x$ is well defined. The discreteness of the particle aspects of quantum  mechanics can be traced back to the discrete occupation numbers that can only take the values zero or one. We also have used discrete time steps, and we have formulated the lattice action such that the Grassmann wave function $g(t+2\epsilon)$ obtains from $g(t)$ by a rotation. The time evolution is unitary not only in the limit $\epsilon\to 0$, but also for finite $\epsilon$. A unitary time evolution for infinitesimal time steps is easily achieved by any antisymmetric evolution generator $K_{\tau\rho}$ for the wave function $q_\tau(t)$,
\be\label{z1}
\partial_tq_\tau(t)=\sum_\rho K_{\tau\rho} q_\rho(t).
\ee
Generating a unitary evolution also for finite smallest time steps imposes restrictions on the form of the lattice action. This, together with the requirement of a real Grassmann action, requires some care for the construction of the model and explains the specific form of the action in comparison with other possible lattice actions.

The proposed evolution equation for the classical statistical ensemble of Ising-spins does not only lead to the Schr\"odinger equation for non-relativistic one-particle states. It entails the full dynamical equations for a quantum field theory of Dirac fermions in an external electromagnetic field. The dynamics of states with an arbitrary number of fermions, including the characteristic interference patterns for indistinguishable fermions in quantum mechanics, is correctly described.

At this point it seems worthwhile to ask some questions about the origin of characteristic features of quantum mechanics in our classical statistical setting. Particle-wave duality is realized by the discreteness of Ising-spins on one side, and the continuous probability distribution or classical wave function on the other side. Interference arises from the formulation of the basic evolution law in terms of the classical wave function. While the classical wave function is real, it can nevertheless take positive and negative values which can add to zero locally. The superposition principle or linearity of the quantum evolution finds a direct origin in the formulation of a dynamical law for the classical statistical ensemble that is linear in the classical wave function. The characteristic physics of phases in quantum mechanics is connected to the presence of a complex structure within the real Grassmann algebra. Planck's constant $\hbar$ appears purely as a conversion factor of units. The uncertainty relations can be obtained directly from the possible solutions of the Schr\"odinger equation.

Finally, one of the most characteristic elements of the mathematical formulation of quantum mechanics is the presence of a non-commutative product for operators and associated observables. These structures are obviously present in our formulation and clearly very useful for a discussion of solutions of the Schr\"odinger equation or Dirac equation. We have not addressed in this paper the fundamental origin of non-commuting operators and refer in this context to related work \cite{GR}, \cite{CWP}, \cite{3A}. An essential point is the observation that a classical statistical ensemble admits many different product structures for observables, and therefore also the definition of many different correlation functions. The correct choice of a correlation function for the description of a sequence of two measurements depends on the details of the measurement process. There exist various idealizations of measurements. 

The classical correlation function (``pointwise multiplication'' $A_\tau B_\tau$) is adapted to a situation of negligible mutual influence between two measurements. It is often not appropriate for the description of measurements in subsystems, that are characteristic for the quantum experiments. For an idealized separation into a subsystem and its environment many different classical observables are grouped into an equivalence class, for which the properties within the subsystem are identical, but the properties relating to the environment differ. Quantum observables are supposed to measure only properties of the subsystem. They are therefore associated to a whole equivalence class of classical observables, rather than to a single classical observable. A one-to-one correspondence between quantum and classical observables is a basic assumption of the Kochen-Specker theorem \cite{KS}, which therefore does not apply in our setting.

The classical product of observables often depends on the properties of the environment and is therefore not suitable for idealized measurements of properties of subsystems. This is the point why Bell's inequalities \cite{Bell}, \cite{BS} do not apply for idealized measurements of properties of subsystems. These inequalities implicitly assume the use of the classical correlation function. In short, Bell's inequalities are circumvented not by abandoning locality or causality, but simply by concentrating on non-classical correlation functions that are appropriate for measurements in subsystems. In other words, the experimental verification that the observed correlations violate Bell's inequalities tells us that the classical correlation function should not be used for this type of measurements. This is in accordance with our arguments that classical correlations do not correspond to idealized measurements of subsystem properties. It has been shown that in many circumstances the idealized measurements of subsystem properties correspond precisely to the quantum correlation function that is associated to the standard non-commutative operator product in quantum mechanics \cite{GR}, \cite{CWP}.

The essence of the emergence of non-commutative structures is the coarse graining of information. Again, this issue has not been addressed in the present paper. It seems reasonable to expect, however, that a system that is governed by the Dirac equation on microphysical scales - say lattice distances shorter than the Planck length - will also show similar properties at ``macroscopic scales'' associated to coarse graining. (Such macroscopic scales can still be much smaller than all characteristic scales of atom physics or elementary particle physics.) The basic reason is that the form of the Dirac equation for one-particle states is essentially fixed by the symmetries. It will not be altered if the coarse graining respects the symmetries. While it remains an interesting task to perform this coarse graining explicitly, the main message of this paper is already very clear at the present stage: Quantum field theory can be obtained from a classical statistical ensemble.

\end{document}